\documentclass[12pt,english]{article}
\usepackage[T1]{fontenc}
\usepackage[latin1]{inputenc}
\usepackage{a4wide}
\usepackage{amsmath}
\usepackage{color}
\usepackage{graphicx}
\usepackage{amssymb}

\usepackage{babel}

\newcommand {\beq}{\begin{equation}}
\newcommand {\eeq}{\end{equation}}
\newcommand {\bea}{\begin{eqnarray}}
\newcommand {\eea}{\end{eqnarray}}
\newcommand {\nn}{\nonumber \\}

\newcommand {\e}{{\rm e}}

\newcommand {\m}{\mu}
\newcommand {\n}{\nu}
\newcommand {\pl}{\partial}

\newcommand {\al}{\alpha}
\newcommand {\be}{\beta}

\newcommand {\Ga}{\Gamma}
\newcommand {\x}{\xi}
\newcommand {\ka}{\kappa}
\newcommand {\la}{\lambda}
\newcommand {\La}{\Lambda}

\newcommand {\om}{\omega}

\newcommand {\ep}{\epsilon}

\newcommand {\na}{\nabla}
\newcommand {\del}  {\delta}
\newcommand {\Del}  {\Delta}
\newcommand {\mn}{{\mu\nu}}

\newcommand {\half}{ {\frac{1}{2}} }

\newcommand {\fourth} {\frac{1}{4} }
\newcommand {\Ecal}{{\cal E}}

\newcommand {\Lcal}{{\cal L}}

\newcommand {\Dcal}{{\cal D}}

\newcommand {\Hcal}{{\cal H}}

\newcommand {\ptil} {{\tilde p}}
\newcommand {\ktil} {{\tilde k}}

\newcommand {\ytil}{{\tilde y}}

\newcommand {\ztil}{{\tilde z}}

\newcommand {\Lhat}{{\hat L}}

\newcommand {\phat}{{\hat p}}

\newcommand {\xhat}{{\hat x}}

\newcommand {\delh} {{\hat \delta}}

\newcommand {\rdot}{\dot{r}}

\newcommand {\Wdot}{\dot{W}}
\newcommand {\xdot}{\dot{x}}
\newcommand {\K}{{\bf K}}
\newcommand {\I}{{\bf I}}

\newcommand {\intfx} {{\int d^4x}}

\newcommand {\intxy} {{\int d^4xdy}}

\newcommand {\intpL} {{\int_{\ptil\leq\Lambda} \frac{d^4p}{(2\pi)^4}}}
\newcommand {\intpE} {{\int \frac{d^4p_E}{(2\pi)^4}}}

\newcommand {\change} {\leftrightarrow}
\newcommand {\ra} {\rightarrow}

\newcommand {\pr}   {{\quad .}}
\newcommand {\com}  {{\quad ,}}
\newcommand {\q}    {\quad}

\newcommand {\qqqqq}   {\quad\quad\quad\quad\quad}

\newcommand {\nl}    {\newline}

\newcommand {\PL}   {Phys.Lett.}
\newcommand {\PR}   {Phys.Rev.}

\newcommand {\PRL}   {Phys.Rev.Lett.}

\newcommand {\JHEP}  {J. High Energy Phys.}

\newcommand {\PTP}  {Prog.Theor.Phys.}

\newcommand {\CQG}  {Class.Quantum.Grav.}

\newcommand {\Pla} {\frac{{\tilde p}}{\omega}}

\newcommand {\Tev} {\frac{{\tilde p}}{T}}

\newcommand {\Hora} {Ho\u{r}ava}      %090615
\begin{document}

\title{Geometric Approach to Quantum Statistical Mechanics and 
Minimal Area Principle
\footnote{
The content was presented (talk in ICSF2010) in \cite{ICSF2010}.
          }
}

\author{S. Ichinose}

\maketitle
\begin{center}\emph{
Laboratory of Physics, School of Food and Nutritional Sciences, 
University of Shizuoka\\
Yada 52-1, Shizuoka 422-8526, Japan
}\end{center}

\begin{abstract}
A geometric approach to 
some quantum statistical systems (including the harmonic oscillator) 
is presented. 
We regard the (N+1)-dimensional Euclidean {\it coordinate} system (X$^i$,$\tau$) 
as the quantum statistical system of N quantum (statistical) variables (X$^i$) and 
one {\it Euclidean time} variable ($\tau$). 
Introducing a path (line or hypersurface) in this space (X$^i$,$\tau$), 
we adopt the path-integral method to quantize the mechanical system. 
%We regard the system variables as the coordinates which 
%describes the N+1 dimensional Euclidean space, 
%N coordinates and one extra coordinate. 
This is a new view of (statistical) quantization of the {\it mechanical} system. 
It is inspired by the {\it extra dimensional model}, appearing in the unified theory of forces 
including gravity, 
using the bulk-boundary configuration. 
The system Hamiltonian appears as the {\it area}. 
We show quantization is realized by  
the {\it minimal area principle} in the present geometric approach. 
When we take a {\it line} as the path, 
the path-integral expressions of the free energy are shown to be 
the ordinary ones (such as N harmonic oscillators) or their simple 
variation. When we take a {\it hyper-surface} as the path, 
the system Hamiltonian is given by the {\it area} of the {\it hyper-surface} 
which is defined as a {\it closed-string configuration} in the bulk space. 
In this case, the system becomes a O(N) non-linear model. 
The two choices,\ (1) the {\it line element} in the bulk ($X^i,\tau $) 
and (2) the Hamiltonian(defined as the damping functional in the path-integral) 
specify the system dynamics. 
%specifies the {\it ensemble type} of the statistical system.
After explaining this new approach, we apply it to a topic in 
the 5 dimensional quantum gravity.  
We present a {\it new standpoint} about the quantum gravity: 
(a)\ The metric (gravitational) field is treated as the background (fixed) 
one;\  
(b)\ The space-time coordinates are not merely position-labels but are quantum 
(statistical) variables by themselves. 
We show the recently-proposed 5 dimensional 
Casimir energy (ArXiv:0801.3064,0812.1263) is valid. 
\end{abstract}

%PACS: 
%PACS NO:
%04.50.+h,\ 
%Gravity in more than four dimensions, Kaluza-Klein theory, 
%unified field theories, alternative theories of
%11.10.Kk,\ 
%Field theories in dimensions other than four 
%11.25.Mj,\ 
%Compactification and four-dimensional models
%12.10.-g 
%Unified field theories and models 
%11.30.Er,\ 
%Charge conjugation,parity,time reversal, and other discrete
%symmetries
%******

\section{Introduction\label{intro}}
%***label**{intro}
%%%%%%%%%%%%%%%%%%%%%%%%%%%%%%%%%%%%%%%%%%%%%%%%%%%%%%%%%%%%%%%%%%%%%%%%%
%%%%%%%%%%%%%%%%%%%%%%%%%%%%%%%%%%%%%%%%%%%%%%%%%%%%%%%%%%%%%%%%%%%%%%%%%
In the quest for the fundamental structure of the space, time, and matter, the most advanced 
theories are the string theory, D-brane theory and M-theory\cite{StringText}. 
They are beyond the quantum field theory in that the extended (in space) 
objects are treated as fundamental elements. Since the finding of AdS/CFT 
correspondence\cite{Malda9711,GKP9802,Witten9802}, various 
new ideas and techniques, developed for them, are imported 
into the {\it non-perturbative} analysis of the quantum field theories. In particular, 
the application to the material physics is marvelous: the heavy ion collision 
physics and the  
viscosity in the quark-gluon plasma(\cite{Natsu0701,Son0704,Mateos0709} 
for review), superconductivity and superfluidity
\cite{Gubser0801,HHH0803,GuPu0805,HHH0810}, 
baryon mass spectrum in QCD\cite{SaSu0412,SaSu0507}. 
In this circumstance, two {\it new standpoints} 
about the space-time quantization 
appear. One is proposed by \Hora\cite{Hora0812,Hora0901}. He introduced Lifshitz's 
higher-derivative scalar theory and its renormalization group behavior into 
his idea about the new quantum gravity. Another one is revively given by E. Verlinde\cite{EVerlinde1001}. 
He emphasizes the entropic force (rather than the energetic force) and the thermodynamical 
behavior near the horizon (Hawking radiation). 
With this recent trend of the geometrical view, the statistical(thermal) view
and the visco-elastic view 
, we present a new formalism where the quantum statistical system is treated 
purely in the geometrical way.     

Let us mention  the present situation of 
the space-time quantization (quantum gravity), 
because it is the concrete motivation of the present work. 
The space-time geometry is specified by the metric tensor field $g_\mn(x)$ 
which appears in the definition of 
the line element $(ds^2)_{4D}=g_\mn(x)dx^\m dx^\n~(\m,\n=0,1,2,3)$. 
One of most important problems of the present theoretical physics is the clarification 
of the {\it quantum role} of the metric (gravitational) field $g_\mn$. We already have 
a long (nearly half century) history of the quantum gravity since Feynman\cite{Fey63} and 
DeWitt \cite{DeW67} pioneered. 
About one decade ago, inspired by the development of the string theory and 
the D-brane theory, a fascinating model of unification of forces 
was proposed. It is a 5 dimensional(dim) model with 
AdS$_5$ geometry and is called "Randall-Sundrum model" or the "warped model"\cite{RS99}. 
This is a representative of the extra dimensional models. 
The most important purpose of the present work is to make this 5 dim model 
{\it legitimate} as the {\it quantum field theory}. 

The AdS$_5$ space-time geometry is described as
%*** intro1B%%%%%%%%%%%%%%%%
\bea
\mbox{Warped Metric (y-expression)}\q
ds^2=\e^{-2\om |y|}\eta_\mn dx^\m dx^\n +dy^2\ ,\ 
-l\leq y\leq l\ ,
\label{intro1B}
\eea 
%%%%%%%%%%%%%%%%%%%%%%%%%%%
where 
$
\{\mu,\nu=0,1,2,3\}\ ,\ 
(\eta_\mn)=\mbox{diag}(-1,1,1,1) 
$. 
$y$ is the extra coordinate. The parameter $\om$ is the 5 dim (bulk) scalar curvature. 
$l$ is the size parameter of the extra coordinate. 
We respect the periodicity: $y\ra y+2l$\ , and Z$_2$-parity: $y\change -y$. 
Instead of $y$, another coordinate $z$ is also used. 
\footnote{
%*** intro1C  ****
$z$ is defined by $y$ as
\bea
z=\left\{
\begin{array}{cc}
\frac{1}{\om}\e^{\om y} &  y>0 \\
0    &  y=0 \\
-\frac{1}{\om}\e^{-\om y} &  y<0 
\end{array}
\right.
\nonumber
%\label{intro1C}
\eea
}
%*** intro1%%%%%%%%%%%%%%%%
\bea
\mbox{Warped Metric (z-expression)}\q
ds^2=\frac{1}{\om^2z^2}(\eta_\mn dx^\m dx^\n+{dz}^2)
=G_{MN}dX^MdX^N \ ,\nn
|z|=\frac{1}{\om}\e^{\om|y|}
\com\q   \frac{1}{\om}<|z|<\frac{1}{T}\com\q
T\equiv \om\e^{-\om l}\com
% G\equiv \det G_{AB}\com
\nn 
R_{MN}=4\om^2G_{MN}
 \com\q R=20\om^2>0
\com\q
\sqrt{-G}=\sqrt{-\det G_{MN}}=\frac{1}{(\om|z|)^5}\com
\label{intro1}
\eea 
%%%%%%%%%%%%%%%%%%%%%%%%%%%
where 
$
(X^M)\equiv (x^\mu,z)\ ,\ \{M,N=0,1,2,3,5\}
$. 
\footnote{
$T$ is {\it not} a temperature parameter but a IR parameter like $l$ ($T=\om\e^{-\om l}$). 
The temperature appears later as $\be^{-1}$. See eq.(\ref{oneHO2}). 
}
The flat (5D Minkowski) limit is obtained by $\om\ra 0$ in the y-expression (\ref{intro1B}). 
%*** intro1b%%%%%%%%%%%%%%%%
\bea
\mbox{Flat Metric}\q
ds^2=\eta_\mn dx^\m dx^\n +dy^2\com\q (X^M)=(x^\m, y)\com\q 
-l\leq y\leq l
\com
\label{intro1b}
\eea 
%%%%%%%%%%%%%%%%%%%%%%%%%%%

\begin{figure}
\caption{
N(=2) dim hypersurface in N+1 dim (Euclidean flat) space $(x^1,x^2,\cdots,x^N,y)=(x^a,y)$. 
Sphere S$^{N-1}$(circles in the figure) at $y$ has the radius $r(y)$. 
%***HySurfF.eps\
        }
\begin{center}
\includegraphics[height=8cm]{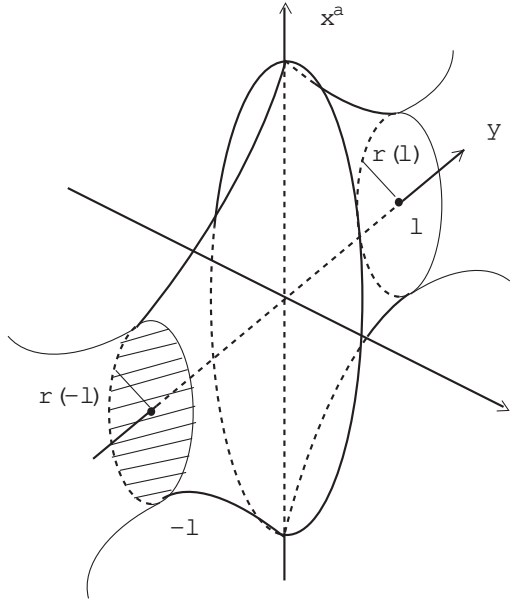}
\end{center}
\label{HySurfF}
\end{figure}

Traditional calculation\cite{AC83,SI0801,SI0812} gives 
the $\La^5$-divergent result for Casimir energy, on the above geometries, 
of 5D models. 
In the calculation, Casimir energy is expressed as 
the 5D space-momentum integral ($\int d^4p_Edy$ or $\int d^4p_Edz$) of 
some energy (density) function $F(\ptil,y)$ or $F(\ptil,z)$. (See Sec.5 for detail.) 
In ref.\cite{SI0801,SI0812}, 
we claim the $\La^5$-divergence comes from this 'naive' integration measure 
and should be replaced by some proper measure, based on close numerical 
calculation using some trial integration measures. 
Finally, Caasimir energy of 
the free fields (electromagnetism, free scalar theory) 
is {\it proposed} to be replaced by the 
following {\it path-integral}. 
%*** intro2 %%%%%%%%%%%%%%%%
\bea
\qqqqq\mbox{For Flat Geometry}\nn
-\Ecal_{Cas}(l,\La)
=\int_{1/\La}^{l}d\rho\int_{r(0)=r(l)=\rho}
\prod_{a,y}\Dcal x^a(y)\left\{\int_0^l F_1(\frac{1}{r(\ytil)},\ytil)
d\ytil\right\}
\mbox{\ }    \nn
\times \exp\left[ 
-\frac{1}{2\al'}\int_{0}^{l}\sqrt{{r'}^2+1}~r^3 dy
    \right],\ r'=\frac{dr}{dy},\nn
\qqqqq\mbox{For Warped Geometry}\nn
-\Ecal_{Cas}(\om,T,\La) 
=\int_{1/\La}^{1/\m}d\rho\int_{r(1/\om)=r(1/T)=\rho}
\prod_{a,z}\Dcal x^a(z)
\left\{\int_{1/\om}^{1/T} F_2(\frac{1}{r(\ztil)},\ztil)d\ztil\right\}\nn
\times\exp\left[ 
-\frac{1}{2\al'}\int_{1/\om}^{1/T}\frac{1}{\om^4z^4}\sqrt{{r'}^2+1}~r^3 dz
    \right]\com\q r'=\frac{dr}{dz}\com
\label{intro2}
\eea 
%%%%%%%%%%%%%%%%%%%%%%%%%%%
where 
$r=\sqrt{\sum_{a=1}^{4}(x^a)^2}$. 
\footnote{
The case $\al'\ra\infty$ in (\ref{intro2}) is essentially the traditional 
definition of Casimir energy. 
}
($\{x^a| a=1,2,3,4\}$ is the {\it Euclideanized} coordinates of 
$\{x^\m |~\mu=0,1,2,3\} $, $x^0=ix^4$.)
In the above proposal, the isotropy of the 4D world $\{x^a\}$ is assumed. 
$F_1$ and $F_2$ are some energy density functions and
will appear later in (\ref{HK20}) and (\ref{HKA11}) respectively. 
$\La$ is the UV-cutoff parameter, 
$\m\equiv\La T/\om$ is the IR-cutoff one and $l$ is the periodicity(IR) one. 
The above path-integrals are over all paths 
of 4 dim {\it hypersurfaces} defined by 
%*** intro3 %%%%%%%%%%%%%%%%
\bea
\mbox{Flat Geometry:}\q\q\sqrt{\sum_{a=1}^{4}(x^a)^2}=r(y)\com\q -l\leq y \leq l\com\nn
\mbox{Warped Geometry:}\q\q\sqrt{\sum_{a=1}^{4}(x^a)^2}=r(z)\com\q \frac{1}{\om}\leq |z| \leq \frac{1}{T}\pr
\label{intro3}
\eea 
%%%%%%%%%%%%%%%%%%%%%%%%%%%
The form of the function $r(y)$ or $r(z)$ specifies the path of the hypersurface. 
See Fig.\ref{HySurfF} for the case of the N+1 dim space. 
                             %%%   <Fig.1   %%%
                              %%%   Fig.1>  %%%
This is a {\it closed string} configurartion. 
The {\it area} (4D volume) plays the role of {\it Hamiltonian} of the 
quantum statistical system $\{x^a\}$. 
$F_i$ comes from the {\it matter-field} quantization and   
plays a role of the {\it energy 'operator'} in the path-integral over the 
4D hyper-surface $r(y)$ or $r(z)$. 
The {\it string (surface) tension} parameter $1/2\al'$ is introduced.  
Note that the proposal (\ref{intro2}) 
is obtained by taking the {\it new standpoint} that 
the (bulk) metric field $G_{MN}(X)$ 
is {\it not} field-quantized and is 
treated as a background field. 
\footnote{
We consider that the form of $G_{MN}(X)$ is given by the 
field equation of the 'effective' action which is obtained 
after the {\it field} quantization of all {\it matter} fields. 
It is a fixed (or background) field in the quantization process of the space-time. 
See also Sec.\ref{Qrole}. 
}
Instead we regard the 
4 dim coordinates $x^a$ as the quantum (statistical) variables, and 
the extra one, $y$ or $z$, as Euclidean time. 
The new point, compared with the 5D Casimir energy calculation so far\cite{AC83}, 
is the introduction of the 'minimal area' factor 
$\exp (-\frac{1}{2\al'}\mbox{Area})=\exp (-\frac{1}{2\al'}\int\sqrt{\det(g_{ab})}d^4x)$ 
where $g_{ab}$ is the {\it induced} metric on the hyper-surface (\ref{intro3}).  
$\al'\ra\infty$ limit, in (\ref{intro2}), goes to the traditional Casimir energy. 
We will show, in this paper, 
the above-type {\it path-integral} very naturally appears in many 
quantum-statistical systems when we view them {\it geometrically}. 
We will show the proposed quantities (\ref{intro2}) are valid. This is the final aim 
of this paper. 

The content is organized as follows. We start with 
the simple quantum statistical system of one harmonic oscillator in Sec.\ref{oneHO}. 
We see the geometric approach 
works well by regarding the {\it extra coordinate} as the {\it Euclidean time}. 
%**compell** us the bulk space 
%be more generalized than the ordinary metric-defined one. 
This approach is shown to give exactly the same result as the ordinary quantization. 
We generalize the harmonic oscillator potential (elastic system) to 
the general one in Sec.3. 
In Sec.4 the one variable system is generalized to the system of N variables.  
We analyze the quantum statistical system in the 
N+1 extra dimensional Euclidean {\it geometry}. 
As the path, we have two choices: line and hypersurface. 
The O(N) {\it nonliner} model 
naturally appears by taking the path of hypersurface which is   
the {\it closed-string} configuration of a special type (\ref{intro3}). 
We stress that taking the area as Hamiltonian is 
one realization of the {\it minimal area principle}. 
In Sec.5, we explain the meaning of the new definition of Casimir energy (\ref{intro2}) 
and present a new treatment of the quantum gravity. 
We conclude in Sec.6. In Appendix A, the content of Sec.4 (N variables elastic system) 
is generalized to the general system.

%%%%%%%%%%%%%%%%%%%%%%%%%%%%  Sec.2  %%%%%%%%%%%%%%%%%%%%%%%%%%%%%%%%%
%%%%  Quantum Statistical System of One Harmonic Oscillator  %%%%%%%%%
%%%%%%%%%%%%%%%%%%%%%%%%%%%%%%%%%%%%%%%%%%%%%%%%%%%%%%%%%%%%%%%%%%%%%%
\section{Quantum Statistical System of Harmonic Oscillator\label{oneHO}}
%***label***{oneHO}

\subsection{'Dirac' Type\label{1HOsup}}
\begin{figure}
\caption{
A path of line in 2D Euclidean space (X,$\tau$). The path starts
at x(0)=$\rho$ and ends at x($\be$)=$\rho'$. 
%***2DPath.eps\
        }
\begin{center}
\includegraphics[height=8cm]{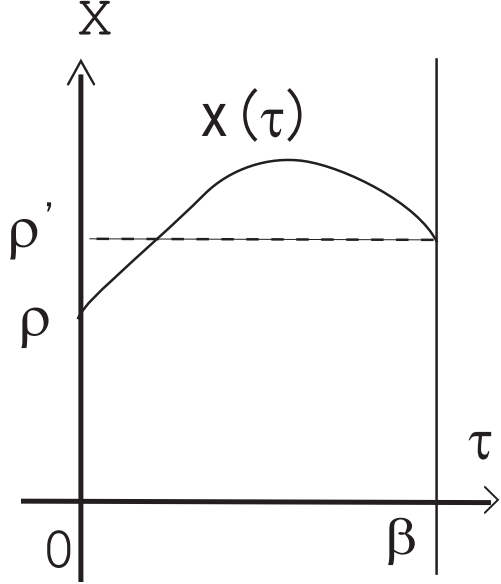}
\end{center}
\label{2DPath}
\end{figure}

%***label***{1HOsup}\nl
Let us consider 2 dim Euclidean space $(X,\tau)$ described by the following metric. 
%*** oneHO1%%%%%%%%%%%%%%%%
\bea
ds^2=dX^2+\om^2X^2d\tau^2=G_{AB}dX^A dX^B\com\nn
(X^A)=(X^1,X^2)=(X,\tau)\com\q (G_{AB})=\mbox{diag}(1,\om^2X^2)\com\nn
R_{AB}=0 \com\q R=G^{AB}R_{AB}=0 \com
\label{oneHO1}
\eea 
%%%%%%%%%%%%%%%%%%%%%%%%%%%
where $A,B=1,2$. $\om$ is the 'spring' constant with the dimension of mass. 
We impose the {\it periodicity} (period: $\beta$) in the direction of the extra dimension $\tau$. 
%*** oneHO2%%%%%%%%%%%%%%%%
\bea
\tau\ra\tau+\be
\pr
\label{oneHO2}
\eea 
%%%%%%%%%%%%%%%%%%%%%%%%%%%
This is a {\it way} to introduce the {\it temperature} ($1/\be$) in the system. 
Here we take a path $\{ x(\tau),\ 0\leq \tau\leq \be\}$ in the 2D bulk space (X,$\tau$) and 
the {\it induced} metric on the {\it line} is given by
%*** oneHO3%%%%%%%%%%%%%%%%
\bea
X=x(\tau)\com\q 
dX=\xdot d\tau\com\q \xdot\equiv\frac{dx}{d\tau}\com\q
0\leq\tau\leq\be\com\nn
ds^2=(\xdot^2+\om^2x^2)d\tau^2
%\left[*********=(1+\frac{\om^2x^2}{\xdot^2})dx^2*******\right]
\pr
\label{oneHO3}
\eea 
%%%%%%%%%%%%%%%%%%%%%%%%%%%
See Fig.\ref{2DPath}. 
                             %%%   <Fig.2   %%%
                              %%%   Fig.2>  %%%

Then the {\it length} L of the path $x(\tau)$ is given by 
%*** oneHO4%%%%%%%%%%%%%%%%
\bea
L=\int ds=\int_0^\beta\sqrt{\xdot^2+\om^2x^2}d\tau
\pr
\label{oneHO4}
\eea 
%%%%%%%%%%%%%%%%%%%%%%%%%%%
We take the half of the length ($\half L$) as the system Hamiltonian 
({\it minimal length principle}). 
Then the free energy $F$ of the system is given by 
\footnote{
When we regard $x$ as the space position (of a particle), the physical 
dimension is that of the length. Then the distribution function 
in (\ref{oneHO5}) is, in general, $\exp (-L/2\al')$ 
where ${\al'}^{-1}$ is the tension parameter with the (length)$^{-1}$ 
dimension. We take $\al'=1$ for simplicity. This note is valid 
for the following some models. 
}
%*** oneHO5%%%%%%%%%%%%%%%%
\bea
\e^{-\be F}=\int_{-\infty}^{\infty}d\rho
\int_{\begin{array}{c}x(0)=\rho\\x(\be)=\rho\end{array}}
\prod_\tau\Dcal x(\tau)\exp \left[-\half\int_0^\beta\sqrt{\xdot^2+\om^2x^2}d\tau\right]
\com
\label{oneHO5}
\eea 
%%%%%%%%%%%%%%%%%%%%%%%%%%%
where the path-integral is done for all possible paths with the indicated boundary condition (b.c.). 
This quantum statistical system can be regarded as the 'fermionic partner' of the ordinary 
harmonic oscillator.
\footnote{
The situation reminds us of the relation between Nambu-Goto action 
and Polyakov action in the string theory\cite{Pol81}. 
The introduction of an auxiliary variable helps to 'normalize' 
the square-root action (\ref{oneHO5}). In this case, the geometric role of 
the auxiliary variable remains obscure.
} 
%*********This is the free energy of the {\it super} harmonic oscillator.  REFERENCES !! **************

\subsection{Standard Type\label{1HOnrm}}
\begin{figure}
\caption{
Singular and regular lines in 2D Euclidean space (X,$\tau$).\ 
(a) regular line, simply increasing;\ 
(b) regular line, maximum at $\beta_{max}$ and minimum at $\beta_{min}$; 
(c) singular line, different derivatives for $\beta\ra\beta_c\pm 0$; 
(d) singular line, divergent at $\beta_{div}$; 
(e) singular line, multi-valued. 
%***2DPaths.eps\
        }
\begin{center}
\includegraphics[height=15cm]{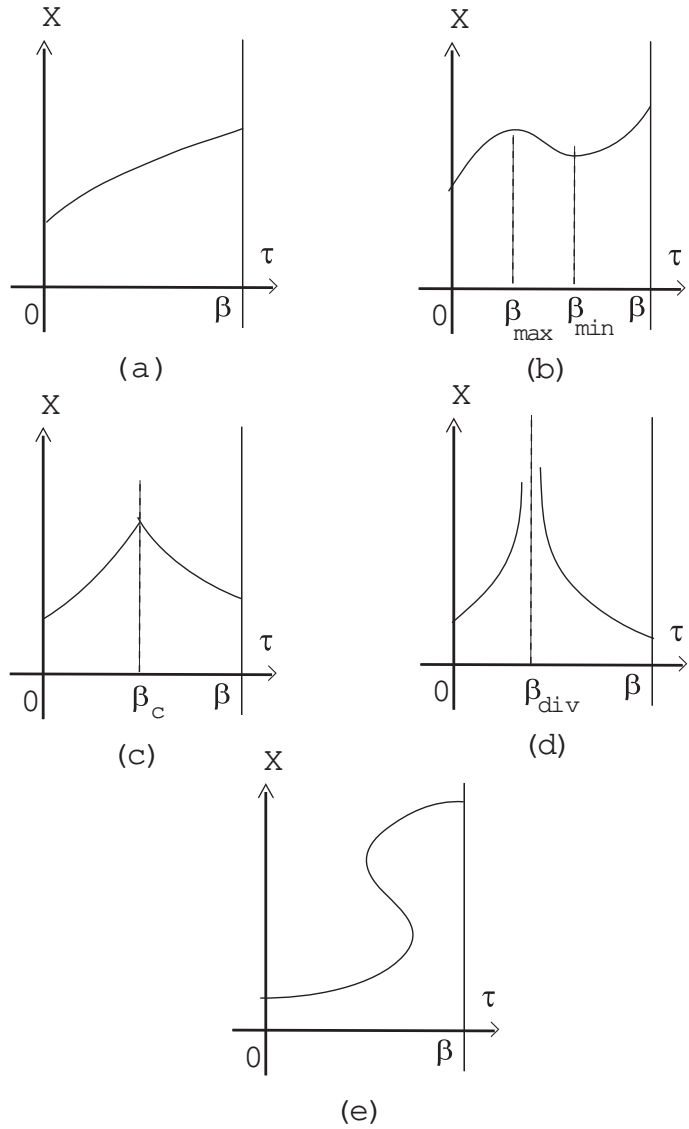}
\end{center}
\label{2DPaths}
\end{figure}

%***label***{1HOnrm}\nl
Now we consider another type of 2 dim Euclidean space $(X,\tau)$ described by the following
{\it line} element. 
%*** oneHO6%%%%%%%%%%%%%%%%
\bea
ds^2=\frac{1}{d\tau^{2}}(dX^2)^2+\om^4X^4d\tau^2+2\om^2X^2dX^2 
=\frac{1}{d\tau^2}(dX^2+\om^2X^2d\tau^2)^2
\com
\label{oneHO6}
\eea 
%%%%%%%%%%%%%%%%%%%%%%%%%%%]
where we put the following condition on the infinitesimal quantities, $d\tau^2$ and $dX^2$,  
in order to keep all terms of (\ref{oneHO6}) in the same order. 
%*** oneHO7%%%%%%%%%%%%%%%%
\bea
\mbox{[Line Element Regularity Condition]}\ :\q\q\q\q\q\q\q\nn  
d\tau^2\sim O(\ep^2)\com\q dX^2\sim O(\ep^2)\com\q
\frac{1}{d\tau^2}dX^2 \sim O(1) 
\com
\label{oneHO7}
\eea 
%%%%%%%%%%%%%%%%%%%%%%%%%%%
where $\ep$ is an arbitrary infinitesimal parameter with the dimension of length. 
\footnote{
The condition (\ref{oneHO7}) 
restricts the trajectory configuration (\ref{oneHO9}) only to 
smooth-lines 
in the 2D bulk space, and excludes singular-lines which have some {\it singular} 
points (the derivative along $\tau$ can not be defined) between $0\leq\tau\leq\be$. 
See Fig.\ref{2DPaths} for singular and regular lines. 
See the final section for the discussions about 
an interpretation of the condition (\ref{oneHO7}). 
}
                             %%%   <Fig.3   %%%
                              %%%   Fig.3>  %%%
Note that we do {\it not} have 2D metric in this case. 
(We {\it cannot} define the bulk metric $G_{AB}(X)$.) 
We impose the {\it periodicity} (period: $\be$).
%*** oneHO8%%%%%%%%%%%%%%%%
\bea
\tau\ra\tau+\be
\pr
\label{oneHO8}
\eea 
%%%%%%%%%%%%%%%%%%%%%%%%%%%
Here we take a path $\{x(\tau),\ 0\leq \tau\leq \be\}$, and 
the {\it induced} metric on the line is given by
%*** oneHO9%%%%%%%%%%%%%%%%
\bea
X=x(\tau)\com\q 
dX=\xdot d\tau\com\q \xdot\equiv\frac{dx}{d\tau}\com\q
0\leq\tau\leq\be\com\nn
ds^2=(\xdot^2+\om^2x^2)^2d\tau^2
%\left[****=(\xdot+\frac{\om^2x^2}{\xdot})^2 dx^2****\right]
\pr
\label{oneHO9}
\eea 
%%%%%%%%%%%%%%%%%%%%%%%%%%%
In the bulk we do {\it not} have the metric, but 
on the path, we {\it do} have this {\it induced} metric. 
Then the {\it length} L of the path $x(\tau)$ is given by 
%*** oneHO10%%%%%%%%%%%%%%%%
\bea
L[x(\tau)]=\int ds=\int_0^\beta(\xdot^2+\om^2x^2)d\tau
\pr
\label{oneHO10}
\eea 
%%%%%%%%%%%%%%%%%%%%%%%%%%%
Hence, taking $\half L$ as the Hamiltonian ({\it minimal length principle}), 
the free energy $F$ of the system is given by 
%*** oneHO11%%%%%%%%%%%%%%%%
\bea
\e^{-\be F}=\int_{-\infty}^{\infty}d\rho
\int_{\begin{array}{c}x(0)=\rho\\x(\be)=\rho\end{array}}
\prod_\tau\Dcal x(\tau)\exp \left[-\half\int_0^\beta(\xdot^2+\om^2x^2) d\tau\right]
\com
\label{oneHO11}
\eea 
%%%%%%%%%%%%%%%%%%%%%%%%%%%
where the path-integral is done for all possible paths with the indicated b.c.. 
This is exactly the free energy of the harmonic oscillator. 
See Feynman's textbook\cite{Fey72}. 
\footnote{
$
F=\frac{\om}{2}+\frac{1}{\be}\ln(1-\e^{-\be\om}), 
E=<\frac{L}{2}>=\frac{\om}{2}\coth(\frac{\om\be}{2})=\frac{\om}{2}+\frac{\om}{\e^{\om\be}-1}, 
S=\frac{1}{T}(E-F)=k\{\frac{\be\om}{2}\coth\frac{\be\om}{2}-\frac{\be\om}{2}-
\ln(1-\e^{-\be\om}) \}
$
}

Note that the condition (\ref{oneHO7}) is necessary for the {\it elastic} 
view to the path. 
%%%%%%%%%%%%%%%%%%%%%%%%%%%%  Sec.3  %%%%%%%%%%%%%%%%%%%%%%%%%%%%%%%%%
%%%%                                                            %%%%%%
%%%%             General Quantum Statistical System             %%%%%%
%%%%                                                            %%%%%%
%%%%%%%%%%%%%%%%%%%%%%%%%%%%%%%%%%%%%%%%%%%%%%%%%%%%%%%%%%%%%%%%%%%%%%
\section{General Quantum Statistical System\label{GQSS}}
%***label***{GQSS}
We generalize the harmonic oscillator potential, $\half \om^2X^2$, to the 
general one $V(X)$.  
As for $V(X)$, we have the following form in mind. 
%*** GQSS0%%%%%%%%%%%%%%%%
\bea
\frac{\om^2}{2}X^2+\frac{\la_3}{3!}X^3+\frac{\la_4}{4!}X^4+\cdots
\com
\label{GQSS0}
\eea 
%%%%%%%%%%%%%%%%%%%%%%%%%%%
where $\la_3, \la_4, \cdots$ are the coupling constants for additional terms. 
%%%%%%%%%%%%%%%%%%%%%%%%%%%
\subsection{'Dirac' Type\label{GQSSsp}}  %***GQSSsp**
We start with the following metric in 2 dim Euclidean space $(X,\tau)$. 
%*** GQSS1%%%%%%%%%%%%%%%%
\bea
ds^2=dX^2+2V(X)d\tau^2=G_{AB}dX^A dX^B\com\nn
(X^A)=(X^1,X^2)=(X,\tau)\com\q (G_{AB})=\mbox{diag}(1,2V(X))\com\nn
(R_{AB})=\left(
\begin{array}{cc}
\frac{V''}{2V}-\fourth\left(\frac{V'}{V}\right)^2,  &  0\\
0\com    &  V''-\half\frac{(V')^2}{V}  
\end{array}
         \right) \com\q 
R=G^{AB}R_{AB}=\frac{V''}{V}-\half\left(\frac{V'}{V}\right)^2 \com\nn
V'\equiv \frac{dV(X)}{dX}\com\q V''\equiv \frac{d^2V(X)}{dX^2}\com
\label{GQSS1}
\eea 
%%%%%%%%%%%%%%%%%%%%%%%%%%%
where $A,B=1,2$. Note that V(X) does {\it not} depend on $\tau$. 
\footnote{
We furthermore note the new standpoint about the quantization of gravity (metric), in the present approach. 
The most familiar way is to regard the 'metric' $V(X)$ as a {\it field} variable at the point ($X,\tau$) 
and quantize it field-theoretically. We do not take such way of quantization. We accept the potential 
form of $V(X)$ as a given one (background treatment) and do not treat $V(X)$ as the quantum 
(field) variable. Instead, we treat the coordinate X as the {\it quantum statistical} 
variable  using the extra coordinate $\tau$ as the Euclidean time. See Sec.\ref{Qrole} furthermore. 
}
We impose the periodicity (period: $\beta$) in the direction of the extra dimension $\tau$ (\ref{oneHO2}). 
On a path $\{ x(\tau),\ 0\leq \tau\leq \be\}$, 
the {\it induced} metric is given by
%*** GQSS2 %%%%%%%%%%%%%%%%
\bea
ds^2=(\xdot^2+2V(x))d\tau^2\com\q 0\leq\tau\leq\be
%\left[*********=(1+\frac{\om^2x^2}{\xdot^2})dx^2*******\right]
\pr
\label{GQSS2}
\eea 
%%%%%%%%%%%%%%%%%%%%%%%%%%%
Hence the {\it length} L of the path $x(\tau)$ is given by 
%*** GQSS3%%%%%%%%%%%%%%%%
\bea
L=\int ds=\int_0^\beta\sqrt{\xdot^2+2V(x)}d\tau
\pr
\label{GQSS3}
\eea 
%%%%%%%%%%%%%%%%%%%%%%%%%%%
Taking the half of the length ($\half L$) as the Hamiltonian, 
we get the free energy $F$ as 
%*** GQSS4%%%%%%%%%%%%%%%%
\bea
\e^{-\be F}=\int_{-\infty}^{\infty}d\rho
\int_{\begin{array}{c}x(0)=\rho\\x(\be)=\rho\end{array}}
\prod_\tau\Dcal x(\tau)\exp \left[-\half\int_0^\beta\sqrt{\xdot^2+2V(x)}d\tau\right]
\pr
\label{GQSS4}
\eea 
%%%%%%%%%%%%%%%%%%%%%%%%%%%

\subsection{Standard Type\label{GQSSn}}
%***label***{GQSSn}\nl
We start with the following line element. 
%*** GQSS5%%%%%%%%%%%%%%%%
\bea
ds^2=\frac{1}{d\tau^{2}}(dX^2)^2+4V(X)^2d\tau^2+4V(X)dX^2 
=\frac{1}{d\tau^{2}}\left(  dX^2+2V(X)d\tau^2\right)^2 
\com
\label{GQSS5}
\eea 
%%%%%%%%%%%%%%%%%%%%%%%%%%%]
where we put the condition (\ref{oneHO7}) on the infinitesimal quantities, $d\tau^2$ and $dX^2$,  
in order to keep all terms in the same order. 
The 2D bulk space do {\it not} have 2D metric. 
We impose the periodicity (period: $\be$) (\ref{oneHO8}).
On a path $\{x(\tau),\ 0\leq \tau\leq \be\}$, we have  
the {\it induced} metric: 
%*** GQSS6 %%%%%%%%%%%%%%%%
\bea
ds^2=(\xdot^2+2V(x))^2d\tau^2
%\left[****=(\xdot+\frac{\om^2x^2}{\xdot})^2 dx^2****\right]
\pr
\label{GQSS6}
\eea 
%%%%%%%%%%%%%%%%%%%%%%%%%%%
The {\it length} L is given by 
%*** GQSS7 %%%%%%%%%%%%%%%%
\bea
L[x(\tau)]=\int ds=\int_0^\beta(\xdot^2+2V(x))d\tau
\pr
\label{GQSS7}
\eea 
%%%%%%%%%%%%%%%%%%%%%%%%%%%
Taking $\half L$ as the Hamiltonian, 
the free energy $F$ is given by 
%*** GQSS8%%%%%%%%%%%%%%%%
\bea
\e^{-\be F}=\int_{-\infty}^{\infty}d\rho
\int_{\begin{array}{c}x(0)=\rho\\x(\be)=\rho\end{array}}
\prod_\tau\Dcal x(\tau)\exp \left[-\half\int_0^\beta(\xdot^2+2V(x)) d\tau\right]
\com
\label{GQSS8}
\eea 
%%%%%%%%%%%%%%%%%%%%%%%%%%%
where the path-integral is done for all possible paths with the indicated b.c..  
This is exactly the free energy of the quantum statistical system of 
one variable $x$ in the general potential V($x$).

%%%%%%%%%%%%%%%%%%%%%%%%%%%%  Sec.4  %%%%%%%%%%%%%%%%%%%%%%%%%%%%%%%%%
%%%%                                                            %%%%%%
%%%%  Quantum Statistical System of N Harmonic Oscillators and  %%%%%%
%%%%                          Nonlinear Model             %%%%%%
%%%%                                                            %%%%%%
%%%%%%%%%%%%%%%%%%%%%%%%%%%%%%%%%%%%%%%%%%%%%%%%%%%%%%%%%%%%%%%%%%%%%%
\section{Quantum Statistical System of N Harmonic Oscillators and O(N) Nonlinear Model\label{NHO}}
%***label***{NHO}

\subsection{'Dirac' Type of N Harmonic Oscillators and 
O(N) nonlinear system\label{NHOsup}}
                             %%%   <Fig.4   %%%
\begin{figure}
\caption{
A path of line $\{x^i(\tau)|i=1,2,\cdots,N\}$ in N(=2)+1 dim space. 
It starts at P=$(\rho_1,\rho_2,\cdots,\rho_N,0)$ and 
ends at P$'$=$(\rho_1',\rho_2',\cdots,\rho_N',\be)$.
%***PathLine.eps\
        }
\begin{center}
\includegraphics[height=8cm]{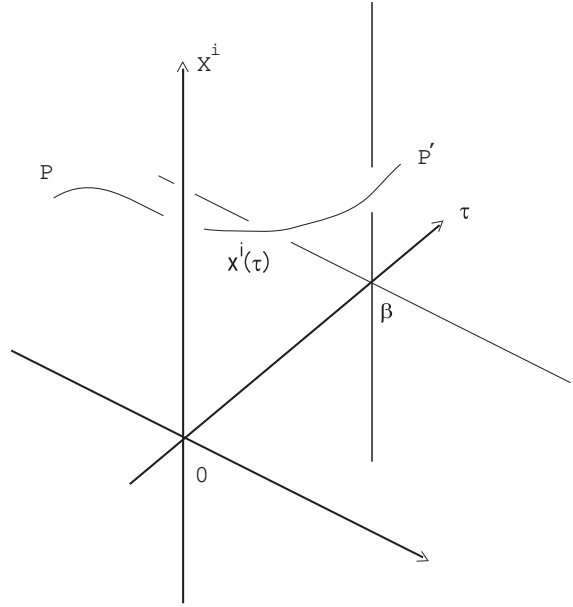}
\end{center}
\label{PathLine}
\end{figure}
                              %%%   Fig.4>  %%%
%***label***{NHOsup}\nl
Let us consider N+1 dim Euclidean space $(X^i,\tau), i=1,2,\cdots ,N$ described by the following metric. 
%*** NHO1%%%%%%%%%%%%%%%%
\bea
ds^2=\sum_{i=1}^N(dX^i)^2+\om^2 d\tau^2\sum_{i=1}^N(X^i)^2=
\sum_{i=1}^N(dX^i)^2+2V(r)d\tau^2=
G_{AB}dX^A dX^B\com\nn
A,B=1,2,\cdots,N,N+1;\q X^{N+1}\equiv \tau\com\q
V(r)=\half\om^2r^2\com \nn
(G_{AB})=\mbox{diag}(1,1,\cdots,1,\om^2r^2)\com\q
r^2\equiv \sum_{i=1}^N(X^i)^2\pr
\label{NHO1}
\eea 
%%%%%%%%%%%%%%%%%%%%%%%%%%%
(Subsec.\ref{1HOsup} is the $N=1$ case. ) 
The Ricci tensor and the scalar curvature are, for N=2, given by
\footnote{
All curvature calculation in this work is checked by 
the algebraic calculation soft "Maxima"\cite{Maxima}. 
}
%*** NHO1b%%%%%%%%%%%%%%%%
\bea
ds^2=dx^2+dy^2+\om^2(x^2+y^2)d\tau^2\com\nn
\left(R_{AB}\right)=\frac{1}{(r^2)^2}
\left(
\begin{array}{ccc}
y^2 & -xy & 0 \\
-yx   & x^2 & 0\\
0    &   0  & \om^2(r^2)^2
\end{array}
\right)
\com\q 
R=\frac{2}{r^2}>0\com\q r^2=x^2+y^2\com\nn
\sqrt{G}=\om\sqrt{x^2+y^2}\com\q
\sqrt{G} R=\frac{2\om}{\sqrt{x^2+y^2}}
\label{NHO1b}
\eea 
%%%%%%%%%%%%%%%%%%%%%%%%%%%
where $(X^1,X^2,X^3)=(x,y,\tau)$ is taken. (See eq.(\ref{GenNHO4}) in App.A, for the general N case using the general potential.)

We impose the periodicity (\ref{oneHO2})(period: $\be$), 
%*** NHO2%%%%%%%%%%%%%%%%
%\bea
%\tau\ra\tau+\be
%\pr
%\label{NHO2}
%\eea 
%%%%%%%%%%%%%%%%%%%%%%%%%%%
and take a path $\{X^i=x^i(\tau)|\ 0\leq \tau\leq \be,\ i=1,2,\cdots,N\}$(See Fig.\ref{PathLine}). 
The {\it induced} metric on the {\it line} is given by
%*** NHO3%%%%%%%%%%%%%%%%
\bea
X^i=x^i(\tau)\com\q 
dX^i=\xdot^i d\tau\com\q \xdot^i\equiv\frac{dx^i}{d\tau}\com\q 
0\leq\tau\leq\be\com
\nn
i=1,2,\cdots,N\com\q
ds^2=\sum_{i=1}^{N}(({\xdot}^i)^2+\om^2(x^i)^2)d\tau^2
\pr
\label{NHO3}
\eea 
%%%%%%%%%%%%%%%%%%%%%%%%%%% 
Then the {\it length} L of the path $\{ x^i(\tau)\}$ is given by 
%*** NHO4%%%%%%%%%%%%%%%%
\bea
L=\int ds=\int_0^\beta\sqrt{ \sum_{i=1}^N((\xdot^i)^2+\om^2{(x^i)}^2) }~d\tau 
\pr
\label{NHO4}
\eea 
%%%%%%%%%%%%%%%%%%%%%%%%%%%
We take the half of the length ($\half L$) as the system Hamiltonian({\it minimal length principle}). 
Then the free energy $F$ of the system is given by 
%*** NHO5%%%%%%%%%%%%%%%%
\bea
\e^{-\be F}=(\prod_i\int_{-\infty}^{\infty}d\rho_i)
\int_{\begin{array}{c}x^i(0)=\rho_i\\x^i(\be)=\rho_i\end{array}}
\prod_{\tau,i}\Dcal x^i(\tau)\exp \left[-\half\int_0^\beta\sqrt{    \sum_{i=1}^N((\xdot^i)^2+\om^2{x^i}^2)}d\tau\right]
\com
\label{NHO5}
\eea 
%%%%%%%%%%%%%%%%%%%%%%%%%%%
where the path-integral is done for all possible paths $\{x^i(\tau);i=1,2,\cdots N\}$ with the indicated b.c.. 
We can regard this as the free energy for 
a variation ('Dirac' type) of the N harmonic oscillators's. 
(See next subsection for the ordinary type of the N harmonic oscillators.)
                             %%%   <Fig.5   %%%
\begin{figure}
\caption{A path of hyper-surface. 
N(=2) dim hypersurface in N+1 dim space $(X^1,X^2,\cdots,X^N,\tau)$. 
S$^{N-1}$ radius $r(\tau)$ starts with $r(0)=\rho$ and ends with $r(\be)=\rho'$. 
We take this configuration as a path in the path integral 
(\ref{NHO9}) and (\ref{NHO18}). This is a closed-string configuration. 
%***PathHySurf.eps\
        }
\begin{center}
\includegraphics[height=8cm]{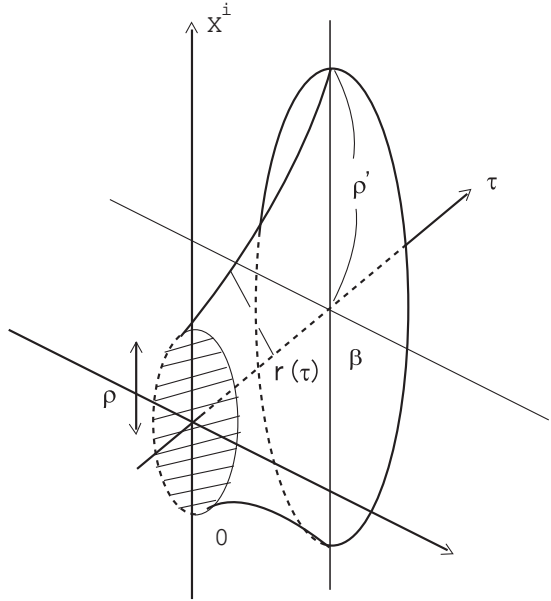}
\end{center}
\label{PathHySurf}
\end{figure}
                              %%%   Fig.5>  %%%

Instead of the length $L$, we can take another geometric quantity. Let us consider 
the following N dim {\it hypersurface} in N+1 dim space (a closed-string configuration). 
See Fig.\ref{PathHySurf} for the N=2 case. 
%*** NHO6%%%%%%%%%%%%%%%%
\bea
\sum_{i=1}^N(X^i)^2=r^2(\tau)\com\q
\sum_{i=1}^NX^idX^i=r\rdot d\tau\com\q 0\leq\tau\leq \be
\pr
\label{NHO6}
\eea 
%%%%%%%%%%%%%%%%%%%%%%%%%%%
The form of 
$r(\tau)$ describes a path (N dimensional hypersurface in the bulk) which is {\it isotropic} in 
the 'brane' at $\tau$ (the N dim plane 'perpendicularly' standing at $\tau$ of the extra axis, 
not the hypersurface ). 
The {\it induced} metric on the N dim hypersurface is given by
%*** NHO7%%%%%%%%%%%%%%%%
\bea
ds^2=\sum_{i,j}(\del_{ij}+\frac{\om^2}{\rdot^2}x^ix^j)dx^idx^j\equiv
\sum_{i,j}g_{ij}dx^idx^j\com\nn
g_{ij}=\del_{ij}+\frac{\om^2}{\rdot^2}x^ix^j\com\q r^2=\sum_{i=1}^{N}(x^i)^2\com\q
\det(g_{ij})=1+\frac{\om^2r^2}{\rdot^2}
\pr
\label{NHO7}
\eea 
%%%%%%%%%%%%%%%%%%%%%%%%%%%
This is the metric of a O(N) 
nonlinear system and is  
the one dimensional {\it nonlinear sigma model} as the field theory.
\footnote{ 
The standard model (2 dim nonlinear sigma model) 
has often been used so far in order to show the {\it renormalization group} behavior of 
various systems. The background (effective action) formulation of the string 
theory heavily relies on the model. 
          }
%
%*cite{4D YM to analogy, be-func, asymp free}, 
%*cite{Polyakov $\be_{ij}$-paper, Wilson RG in the coupling space, 
%Friedan 2D nonlinear sigma model 85}. 
Then the {\it area} of the N dim hypersurface, $A_N$, is given by 
%*** NHO8%%%%%%%%%%%%%%%%
\bea
A_N=\int\sqrt{\det g_{ij}}~d^Nx=\frac{N\pi^{N/2}}{\Ga(\frac{N}{2}+1)}\int\sqrt{\rdot^2+\om^2 r^2}r^{N-1}d\tau
\pr
\label{NHO8}
\eea 
%%%%%%%%%%%%%%%%%%%%%%%%%%%
When we take $\half A_N$ as the Hamiltonian ({\it minimal area principle}), 
the free energy $F$ is given by 
%*** NHO9%%%%%%%%%%%%%%%%
\bea
\e^{-\be F}=\int_{0}^{\infty}d\rho
\int_{\begin{array}{c}r(0)=\rho\\r(\be)=\rho\end{array}}
\prod_{\tau,i}\Dcal x^i(\tau)\exp \left[
-\half\frac{N\pi^{N/2}}{\Ga(\frac{N}{2}+1)}
\int\sqrt{\rdot^2+\om^2 r^2}r^{N-1}d\tau
                                  \right]
\pr
\label{NHO9}
\eea 
%%%%%%%%%%%%%%%%%%%%%%%%%%%
We should compare this result ($N=4$) with the proposed 5D Casimir energy for the {\it flat} 
geometry (\ref{intro2}). 
The component $\sqrt{\rdot^2+\om^2r^2}$ in the integrand of (\ref{NHO9}) 
is replaced by $\sqrt{\rdot^2+1}$ in (\ref{intro2}).  

We recognize, if we start with 
%*** NHO9c%%%%%%%%%%%%%%%%
\bea
ds^2=\sum_{i=1}^{N}(dX^i)^2+d\tau^2\q (\mbox{N+1 dim Euclidean flat})\com
\label{NHO9c}
\eea 
%%%%%%%%%%%%%%%%%%%%%%%%%%%
instead of (\ref{NHO1}), the integration measure becomes {\it exactly} the same as (\ref{intro2}). 
\footnote{
The starting line element (\ref{NHO1}) can be written as a general form: 
$ds^2=\sum_{i=1}^N(dX^i)^2+2V(r)d\tau^2$. 
The content of this subsection is valid for this general potential $V(r)$. 
(\ref{NHO9c}) is the case $V=1/2$. See App.A. 
}

\subsection{Standard Type of N Harmonic Oscillators\label{NHOnrm}}
%***label***{NHOnrm}\nl
Now we consider another type of N+1 dim Euclidean space $(X^i,\tau);\ i=1,2,\cdots N$ described by the following
line element. 
%*** NHO9b%%%%%%%%%%%%%%%%
\bea
ds^2=d\tau^{-2}\{\sum_{i=1}^N(dX^i)^2\}^2+\om^4\{\sum_{i=1}^N(X^i)^2\}^2d\tau^2
+2\om^2\{\sum_{i=1}^N(X^i)^2\} \{\sum_{j=1}^N(dX^j)^2\}  \nn
=\frac{1}{d\tau^{2}}\{ 
\sum_{i=1}^N(dX^i)^2+2V(r)d\tau^2
                     \}^2  
\com\q V(r)=\frac{\om^2}{2}r^2\com\q r^2=\sum_{i=1}^N(X^i)^2\com
\label{NHO9b}
\eea 
%%%%%%%%%%%%%%%%%%%%%%%%%%%]
with the condition:  
%*** NHO10%%%%%%%%%%%%%%%%
\bea
\mbox{[Line Element Regularity Condition]}\ :\q\q\q\q\q\q\nn  
d\tau^2\sim O(\ep^2)\com\q (dX^i)^2\sim O(\ep^2)\com\q
\frac{1}{d\tau^2}\{\sum_{i=1}^N(dX^i)^2\}\sim O(1)
\ ,\ 
\label{NHO10}
\eea 
%%%%%%%%%%%%%%%%%%%%%%%%%%%
in order to keep all terms of (\ref{NHO9b}) in the order of $\ep^2$. 
\footnote{
As in (\ref{oneHO7}), this condition restricts the trajectory configuration 
(\ref{NHO12}) only to {\it smooth} hyper-surfaces in the (N+1)-dim space. 
}
Again we 
note that, in the above case, we do {\it not} have N+1 dim (bulk) metric. 
We impose the periodicity (\ref{oneHO2}): (period: $\be$).
%*** NHO11%%%%%%%%%%%%%%%%
%\bea
%\tau\ra\tau+\be
%\pr
%\label{NHO11}
%\eea 
%%%%%%%%%%%%%%%%%%%%%%%%%%%

Here we take a path of Fig.\ref{PathLine}:\ $\{x^i(\tau)|\ 0\leq \tau\leq \be, i=1,2,\cdots,N\}$ and 
the {\it induced} metric on the path is given by
%*** NHO12%%%%%%%%%%%%%%%%
\bea
X^i=x^i(\tau)\com\q dX^i=\xdot^i d\tau\com\q \xdot^i\equiv\frac{dx^i}{d\tau}\com\q 
0\leq\tau\leq\be\com\nn
ds^2=[\sum_{i=1}^N((\xdot^i)^2+\om^2(x^i)^2)]^2 d\tau^2
\pr
\label{NHO12}
\eea 
%%%%%%%%%%%%%%%%%%%%%%%%%%%
Then the {\it length} L of the path $\{x^i(\tau)\}$ is given by 
%*** NHO13%%%%%%%%%%%%%%%%
\bea
L[x^i(\tau)]=\int ds=\int_0^\beta\sum_{i=1}^N((\xdot^i)^2+\om^2(x^i)^2)d\tau
\pr
\label{NHO13}
\eea 
%%%%%%%%%%%%%%%%%%%%%%%%%%%
Hence, taking $\half L$ as the Hamiltonian ({\it minimal length principle}
), 
the free energy $F$ of the system is given by 
%*** NHO14%%%%%%%%%%%%%%%%
\bea
\e^{-\be F}=\left( \prod_i\int_{-\infty}^{\infty}d\rho_i \right)
\int_{\begin{array}{c}x^i(0)=\rho_i\\x^i(\be)=\rho_i\end{array}}
\prod_{i,\tau}\Dcal x^i(\tau)\exp \left[-\half\int_0^\beta\sum_{i=1}^N((\xdot^i)^2+\om^2(x^i)^2)d\tau
                                  \right]
,
\label{NHO14}
\eea 
%%%%%%%%%%%%%%%%%%%%%%%%%%%
where the path-integral is done for all possible paths with the indicated b.c.. This 
is {\it exactly} the free energy 
of N harmonic oscillators. 

We note again the condition (\ref{NHO10}) is necessary for the {\it elastic} 
view to the hyper-surface. 

\subsection{Middle type of O(N) nonlinear system\label{NBG}}
%***label***{NBG}\nl
Instead of (\ref{NHO9b}), we can start from a slightly modified metric.   
%*** NHO15%%%%%%%%%%%%%%%%
\bea
ds^2=\om^4\{\sum_{i=1}^N(X^i)^2\}^2d\tau^2
+2\om^2\ka\{\sum_{i=1}^N(X^i)^2\} \{\sum_{j=1}^N(dX^j)^2\}  \nn
=\om^2r^2\left( \om^2r^2d\tau^2+2\ka\sum_{j=1}^N(dX^j)^2\right)
=4V(r)\left( V(r)d\tau^2+\ka\sum_{j=1}^N(dX^j)^2 \right)         \com\nn 
V(r)=\frac{\om^2}{2}r^2\com\q
r^2=\sum_{i=1}^N(X^i)^2
\pr
\label{NHO15}
\eea 
%%%%%%%%%%%%%%%%%%%%%%%%%%%
We drop the first term of (\ref{NHO9b}), and add 
a free (real) parameter $\ka$ in the third one. 
We stress that, in this case, we need {\it not} the condition of (\ref{NHO10}). 
The line element is the ordinary one and 
we have the bulk metric $G_{AB}$ in this case. 
The Ricci tensor and the scalar curvature, for $N=2$, are given by 
%*** NHO15b%%%%%%%%%%%%%%%%
\bea
ds^2=\om^4(x^2+y^2)^2d\tau^2+2\om^2\ka (x^2+y^2)(dx^2+dy^2)\com\nn
\left(R_{AB}\right)=\frac{1}{(r^2)^2}
\left(
\begin{array}{ccc}
4y^2 & -4xy & 0 \\
-4yx   & 4x^2 & 0\\
0    &   0  & \frac{2\om^2}{\ka}(r^2)^2
\end{array}
\right)
\com\q 
R=\frac{4}{\ka\om^2(r^2)^2}\com\q r^2=x^2+y^2\com\nn
\sqrt{G}=2\om^4 |\ka| r^4\com\q \sqrt{G}R=8\om^2\cdot\mbox{sign}(\ka)\com
\label{NHO15b}
\eea 
%%%%%%%%%%%%%%%%%%%%%%%%%%%
where $(X^1,X^2,X^3)=(x,y,\tau)$ and sign$(\ka)$ is the sign of $\ka$.
\footnote{
$R>0 \q\mbox{for}\q \ka>0\com\q R<0 \q\mbox{for}\q \ka<0$. 
} 
(See eq.(\ref{GenNHO15b}) in App.A for general potential.) 
We consider the N dim hypersurface (\ref{NHO6}), or Fig.\ref{PathHySurf}, 
and the {\it induced} metric on it is 
given by 
%*** NHO16%%%%%%%%%%%%%%%%
\bea
ds^2=\sum_{i,j=1}^N 2\om^2r^2(\ka\del_{ij}+\half\frac{\om^2}{\rdot^2}x^ix^j)dx^idx^j\equiv
\sum_{i,j}g_{ij}dx^idx^j
\pr
\label{NHO16}
\eea 
%%%%%%%%%%%%%%%%%%%%%%%%%%%
Then the {\it area} of this hypersurface is given by 
%*** NHO17%%%%%%%%%%%%%%%%
\bea
A_N=\int\sqrt{\det g_{ij}}~d^Nx=
\frac{(2\pi\om^2|\ka|)^{N/2}}{\Ga(\frac{N}{2}+1)}
\int_0^\be r^N\sqrt{\rdot^2+\frac{r^2\om^2}{2|\ka|}}~r^{N-1}d\tau
\pr
\label{NHO17}
\eea 
%%%%%%%%%%%%%%%%%%%%%%%%%%%
Taking $\half A_N$ as the Hamiltonian ({\it minimal area principle}), 
the free energy, $F$, is given by 
%*** NHO18%%%%%%%%%%%%%%%%
\bea
\e^{-\be F}=\int_{0}^{\infty}d\rho
\int_{\begin{array}{c}r(0)=\rho\\r(\be)=\rho\end{array}}
\prod_{\tau,i}\Dcal x^i(\tau)\exp \left[
-\half 
\frac{(2\pi\om^2|\ka|)^{N/2}}{\Ga(\frac{N}{2}+1)}
\int_0^\be r^N\sqrt{\rdot^2+\frac{r^2\om^2}{2|\ka|}}~r^{N-1}d\tau
                                  \right]
\pr
\label{NHO18}
\eea 
%%%%%%%%%%%%%%%%%%%%%%%%%%%
We should compare this result (N=4,$\ka$=1/2) with the proposed 5D Casimir energy for 
the {\it warped} geometry (\ref{intro2}). They are similar 
(
$(\om r)^4\sqrt{\rdot^2+r^2\om^2}$ of (\ref{NHO18}) is replaced by 
$(1/\om z)^4\sqrt{{r'}^2+1}$ of (\ref{intro2}). 
). The exactly same one is obtained in the next subsection.

\subsection{Modified type of O(N) nonlinear system\label{Mod}}
%***label***{Mod}\nl
Instead of (\ref{NHO15}), we take the following modified type metric.   
%*** NHO19b %%%%%%%%%%%%%%%%
\bea
ds^2=
W(\tau)\left( 2 V(r)d\tau^2+\sum_{j=1}^N(dX^j)^2\right)\com\q r^2=\sum_{i=1}^N(X^i)^2
\pr
\label{NHO19b}
\eea 
%%%%%%%%%%%%%%%%%%%%%%%%%%%
We recognize, if we start with $W(\tau)=\frac{1}{\tau^2},\ V(r)=\half$: 
%*** NHO19%%%%%%%%%%%%%%%%
\bea
\mbox{Euclidean }(\mbox{AdS})_{N+1}:\q ds^2=\frac{1}{\tau^2}\{ d\tau^2+\sum_{j=1}^N(dX^j)^2\}  
\com
\label{NHO19}
\eea 
%%%%%%%%%%%%%%%%%%%%%%%%%%%
instead of (\ref{NHO15}), the integration measure {\it exactly} becomes 
the same as the warped case in (\ref{intro2}):  
$\tau^{-N}\sqrt{\rdot^2+1}r^{N-1}d\tau$.\nl 
$\mbox{}$

The content in this section is generalized for the {\it general} isotropic potential in App.A.

%%%%%%%%%%%%%%%%%%%%%%%%%%%%  Sec.5  %%%%%%%%%%%%%%%%%%%%%%%%%%%%%%%%%
%%%%                                                            %%%%%%
%%%%  Quantum Role of Space-Time Coordinates and                %%%%%%
%%%%  the Matter Fields  - New Treatment of Quantum Gravity     %%%%%%
%%%%                                                            %%%%%%
%%%%%%%%%%%%%%%%%%%%%%%%%%%%%%%%%%%%%%%%%%%%%%%%%%%%%%%%%%%%%%%%%%%%%%
\section{Quantum Role of Space-Time Coordinates and 
the Matter Fields  - New Treatment of Quantum Gravity -\label{Qrole}}
%***label***{Qrole}

The present standpoint on the metric field $G_{MN}(X)$ is that 
it is {\it not} the quantum-field variable and the form (the dependency on $X$) 
is {\it not} affected in the field-quantization process. All other fields (, other than 
the metric field, such as the electromagnetic fields, scalar fields, the 
fermion fields, the gluon fields, etc.)
\footnote{
We call them matter fields.
}
 are treated as the quantum-field variables. 
\footnote{
The metric field is differently treated from the matter fields (all other fields) in the field-quantization process. 
}
The metric field plays only the role of the {\it background} or {\it fixed} field. 
The form of $G_{MN}(X)$ is, in principle, given by the solution of the field equation of 
the 'effective' action which is obtained after the {\it field}-quantization of 
all {\it matter} fields. 
The quantum behavior of the space-time 
is realized as the {\it statistical mechanics} of the coordinates $X^M$ as 
described in the previous sections. 
%%%%%%%%%%%%%%%%%%%%%%%%%%%
%Traditional calculation\cite{AC83,SI0801,SI0812} gives 
%the $\La^5$-divergent result for Casimir energy of the above models. 
%The {\it closed} expression of the energy has recently been presented 
%both for the flat\cite{SI0801} and for the warped\cite{SI0812} cases as follows.
 
The {\it traditional} definition of Casimir Energy of the 5D electromagnetic field 
theory is, for the flat case (\ref{intro1b}), 
%*** HK2%%%%%%%%%%%%%%%%
\bea
\e^{-l^4E_{Cas}}=\left.\int\Dcal A \exp\left[
i\intxy(\Lcal^{5D}_{EM}+\Lcal_{gauge})
                                       \right]
                   \right|_{\mbox{Euclid}}\com\hspace{40mm}\nn
\Lcal^{5D}_{EM}[A_M(X)]=-\fourth F_{MN}F^{MN}\ ,\ 
F_{MN}=\pl_MA_N-\pl_NA_M\ ,\ 
\Lcal_{gauge}[A_M(X)]=-\half(\pl_MA^M)^2\ .                   
\label{HK2} 
\eea
%%%%%%%%%%%%%%%%%%%%%%%%%%%%%
The expression of $E_{Cas}$ defined above, is given by \cite{SI0801}
%*** HK20%%%%%%%%%%%%%%%%
\bea
\hspace{50mm}\mbox{For Flat Geometry (5 dim elctromagnetism) :}\nn
E_{Cas}(l)=\intpL\int_0^ldy (F_f^-(\ptil,y)+4F_f^+(\ptil,y))\com\nn
F_f^\mp(\ptil,y)
=-\int_\ptil^\infty d\ktil\frac{\mp\cosh\ktil(2y-l)+\cosh\ktil l}{2\sinh(\ktil l)}
\pr
\label{HK20}
\eea
%%%%%%%%%%%%%%%%%%%%%%%%%%%%%
The plus-minus symbol, $\mp$, indicates the contribution from 
Z$_2$-parity odd (-) and even (+) components. 
$\ptil$ is the maginitude of 4D momentum $(p_a)=(p_1,p_2,p_3,p_4)$. 
The coincidence with the previous result\cite{AC83} was confirmed\cite{SI0801}. 
As for the warped case (\ref{intro1}), the {\it traditional} definition, for the 5D free scalar theory, is 
given by\cite{SI0812}
%*** KKexp1b%%%%%%%%%%%%%%%%
\bea
\e^{-T^{-4}E_{Cas}}=\left.\int\Dcal\Phi~\exp
\left[
i\int d^5X\sqrt{-G}\Lcal^{5D}_s
\right]              \right|_{\mbox{Euclid}}\nn
=\int\Dcal\Phi(X)\exp\left[
\intfx dz\frac{1}{(\om z)^5}\half\Phi\{
\om^2z^2\pl_a\pl^a\Phi+(\om z)^5 \Lhat_z \Phi
                                       \} 
                     \right]\com\nn
\Lcal^{5D}_s[\Phi(X);X]=-\half \na^M\Phi\na_M\Phi-\half m^2\Phi^2\com\nn
\frac{1}{\om}<|z|<\frac{1}{T}\com\q
\Lhat_z=\frac{d}{dz}\frac{1}{(\om z)^3}\frac{d}{dz}-\frac{m^2}{(\om z)^5}
\com\q (m^2=-4\om^2)
\pr
\label{KKexp1b}
\eea 
%%%%%%%%%%%%%%%%%%%%%%%%%%%
where 
$\Lhat_z$ is the kinetic operator in the extra space (Bessel differential operator). 
Casimir energy $E_{Cas}$ defined in (\ref{KKexp1b}) is explicitly given by 
%*** HKA11%%%%%%%%%%%%%%%%
\bea
\hspace{50mm}\mbox{For Warped Geometry (5 dim Free Scalar, $m^2=-4\om^2$):}\nn
-E^\mp_{Cas}(\om,T)
=\left.\intpE\right|_{\ptil\leq\La}\int_{1/\om}^{1/T}dz~F_w^\mp(\ptil,z)\com 
\q
F_w^\mp(\ptil,z)= \frac{1}{(\om z)^3}\int_{\ptil^2}^\infty\{G_k^\mp (z,z)\}dk^2\com\nn
G_p^\mp(z,z')=\mp\frac{\om^3}{2}z^2{z'}^2
\frac{\{\I_0(\Pla)\K_0(\ptil z)\mp\K_0(\Pla)\I_0(\ptil z)\}  
      \{\I_0(\Tev)\K_0(\ptil z')\mp\K_0(\Tev)\I_0(\ptil z')\}
     }{\I_0(\Tev)\K_0(\Pla)-\K_0(\Tev)\I_0(\Pla)},\nn     
(\Lhat_z-p^2s(z))G^{\mp}_p(z,z')=
\left\{
\begin{array}{ll}
\ep(z)\ep(z')\delh (|z|-|z'|) & \mbox{for\ \ P=}-1 \\
\delh (|z|-|z'|) & \mbox{for\ \ P=}1 
\end{array}
        \right.
\com\q s(z)=\frac{1}{(\om z)^3}\com         
\label{HKA11}
\eea
%%%%%%%%%%%%%%%%%%%%%%%%%%%%%
where $\I_0$ and $\K_0$ are the modified Bessel functions of 0-th order. 
%Both in (\ref{HK20}) and in (\ref{KKexp1b}), Casimir energy is expressed as 
%the 5D space-momentum integral ($\int d^4p_Edy$ or $\int d^4p_Edz$) of 
%energy (density) $F(\ptil,y)$ or $F(\ptil,z)$. 

Casimir energy defined above, 
which has been traditionally calculated, gives $\La^5$-divergence. 
The integral $\intpE dz$ ($\int\frac{d^4p}{(2\pi)^4}dy$) appearing in 
eq.(\ref{HKA11})\ ((\ref{HK20})) corresponds to 
the summation over all positions in 5 dim bulk space $\int d^4xdz$ ($\int d^4xdy$). 
The above expression says $E_{Cas}$ is the total sum of $F(r^{-1},z)$ ($F(r^{-1},y)$) over 
the bulk space positions. We notice here the $\La^5$ divergence 
comes from the fact that 
we have overlooked some proper {\it integration measure}. 
The summation, or the {\it averaging} procedure (of F) should be properly 
defined at this stage. In the present standpoint 
we regard the {\it coordinate} system $(x^a,z)$ ($(x^a,y)$) as the {\it quantum statistical} system 
and consider 
that the coordinate $x^a$ is 
the {\it quantum mechanical} variable with the extra one $z$ ($y$) as {\it Euclidean time}. The traditional  
treatment (simple summation over the set of positions) should 
be corrected by the present quantum (geometric) approach. 
We have proposed it should be done by 
the {\it path-integral} over all hypersurfaces in the bulk space ($x^a,z$) (($x^a,y$)), as described 
in the previous sections. Hence 
the right expression of Casimir energy is given by (\ref{intro2}). 

%In ref.\cite{SI0801,SI0812}, 
%we claim the $\La^5$-divergence comes from this 'naive' integration measure. 
%and should be replaced by some proper measure, based on close numerical 
%calculation using some trial integration measures. 

%%%%%%%%%%%%%%%%%%%%%%%%%%%%  Sec.6  %%%%%%%%%%%%%%%%%%%%%%%%%%%%%%%%%
%%%%                                                            %%%%%%
%%%%  Discussion and Conclusion                                 %%%%%%
%%%%                                                            %%%%%%
%%%%%%%%%%%%%%%%%%%%%%%%%%%%%%%%%%%%%%%%%%%%%%%%%%%%%%%%%%%%%%%%%%%%%%
\section{Discussion and Conclusion\label{conc}}
%***label***{conc}

We have shown some quantum statistical systems of N variables can be 
described by the path (line or hypersurface) integral over the N+1 dim Euclidean space 
with an appropriate Hamiltonian ({\it length} of the line or {\it area} of 
the hypersurface). The system dynamics is determined by choosing the following 
two things:\ 
1)\ With which bulk metric does one start and 2)\ which type of path (line or hypersurface) 
does one take.  
The choice 1) specifies the bulk geometry and 
the choice 2) specifies the {\it embedded geometry} of the path. 
This is the {\it geometric view} of the quantum statistical system. The result is applied to 
Casimir energy of 5 dim models and we show the proposed new definition (\ref{intro2}) is valid. 

As shown in (\ref{NHO9c}) and (\ref{NHO19}), the bulk metrics for 
the integration measures (\ref{intro2}) are standard. Hence the conditions (\ref{oneHO7}) and (\ref{NHO10}) are {\it not} 
necessary only for the proof (of the correctness of (\ref{intro2})). 
But the conditions are important for the {\it elastic} (Harmonic oscillator) 
view of the hyper-surface (See (\ref{oneHO11}) and (\ref{NHO14})). 
More generally they are important when we view the hyper-surface 
as the {\it quantum mechanical system} of the potential $V(x)$ 
(See (\ref{GQSS8}) and (\ref{GenNHO14})).  
 Hence, in this last paragraph, 
we argue the meaning of the line element regularity condition (\ref{oneHO7}) or 
(\ref{NHO10}). 
Traditionally  
the quantum mechanics is formulated by using the operators $\{\xhat^i,\phat^i|i=1,2,\cdots,N\}$ 
which satisfy {\it Heisenberg algebra}:
%*** conc0 %%%%%%%%%%%%%%%%
\bea
[ \xhat^i\ , \phat^j] = i\hbar \del_{ij}
\pr
\label{conc0}
\eea 
%%%%%%%%%%%%%%%%%%%%%%%%%%%
where $\phat^i$ is the momentum operator conjugate of $\xhat^i$. 
This quantum system is characterized by 
the {\it uncertainty relation} 
among the expectation values of these operators 
: 
%*** conc0b %%%%%%%%%%%%%%%%
\bea
\Del x^i\Del p^j \ge \half\hbar~\del_{ij}
\pr
\label{conc0b}
\eea 
%%%%%%%%%%%%%%%%%%%%%%%%%%%
In relation to this uncertainty equation, let us discuss a possible meaning of the 
equations of (\ref{oneHO7}) and (\ref{NHO10}). 
%If we want the {\it  elastic} (harmonic oscillator) view to the hypersurface, 
%we need the condition. 
They guarantee the smooth surface (differentiable 
in the extra-coordinate direction). 
In this case, the metric does {\it not} exist in the bulk, instead we have 
the {\it induced metric} on the path (hypersurface). 
Although the geometric formulation is done in the N+1 dim bulk space, 
we can regard only the hypersurface as the 
'ordinary' world in the sense that the metric is defined only on the hypersurface.
 
\begin{figure}
\caption{
Sphere lattice in the 4D Euclidean momentum space ($p_1,p_2,p_3,p_4$). 
The big 4D ball (radius $\La$) is composed of many small
4D balls (radius $\mu$). The density of small ones shows the 'resolution' 
of this regularization of the 4D continuous manifold. 
%***SphereLattice.eps\
        }
\begin{center}
\includegraphics[height=4cm]{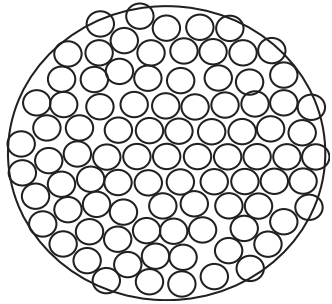}
\end{center}
\label{SphereLattice}
\end{figure}

In the 'discrete' or 'regularized' level, we can take the following 
configuration as the 4D plane 'perpendicularly' standing at a fixed extra-axis 
($z$ or $\tau$) point.
We have the integral over 5D space, $\int d^4p_E\int dz$, in the expression of 
the AdS$_5$ Casimir energy (\ref{HKA11}). The 4D integral part, $\int d^4p_E$, can be 
regularized as
%*** conc0c %%%%%%%%%%%%%%%%
\bea
\mu\leq\sqrt{{p_E}^2}\leq \La\com\q \om^{-1}<z<T^{-1}\com\nn
\mu :\ \mbox{IR cutoff}\com\q \La :\ \mbox{UV cutoff}\com\q
\om :\ \mbox{5D bulk curvature}\com\nn 
T=\om\e^{-l\om}\ \ (l :\ \mbox{periodicity})
\com
\label{conc0c}
\eea 
%%%%%%%%%%%%%%%%%%%%%%%%%%%
where ${p_E}^2\equiv {p_1}^2+{p_2}^2+{p_3}^2+{p_4}^2,\ p_4=i~p_0$. 
The finite range of $z$ comes from the fact taht the AdS$_5$ geometry 
(bulk curvature $\om$) is the concerned manifold and we take the periodic b.c. 
(periodicity $l$) w.r.t. the extra axis. The restricted range, $\mu\leq\sqrt{{p_E}^2}\leq\La$, 
for the Euclidean 4D momentum ($p^a$) comes from the regularization 
of this continuous 4D manifold ('Brane'). We can express this regularization 
as the sphere-lattice shown in Fig.\ref{SphereLattice}. 
                             %%%   <Fig.6   %%%
                              %%%   Fig.6>  %%%

As the regularization, usually $\mu$ and $\La$ are taken independent 
as far as the following condition:\ 
%*** conc0d %%%%%%%%%%%%%%%%
\bea
\mu\q\ll\q\La
\com
\label{conc0d}
\eea 
%%%%%%%%%%%%%%%%%%%%%%%%%%%
is satisfied. 
In the present treatment, however, we take a special regularization by imposing 
the equality between $\frac{\La}{\mu}$ and $\frac{\om}{T}$.\cite{RS01} 
%*** conc0e %%%%%%%%%%%%%%%%
\bea
\mbox{IR-UV harmonic relation}\ :\q\frac{\La}{\mu}=\frac{\om}{T}\gg 1\pr
\label{conc0e}
\eea 
%%%%%%%%%%%%%%%%%%%%%%%%%%%
We call this equality condition 
"IR-UV harmonic relation". This relation helps us to regularize 
both 4D manifold and the extra world in the 'harmonious' way. 
This is a simple way to reduce the number of independent regularization parameters. 

\begin{figure}
\caption{
Eqs.(\ref{conc2}) and (\ref{conc3}) suggest the foam-like structure in the bulk space ($X^i, \tau$). 
Each "cell" has the size $\al_\tau~\times~\al_X$ approximately in the length units shown 
in the figure. 
%***foam.eps\
        }
\begin{center}
\includegraphics[height=4cm]{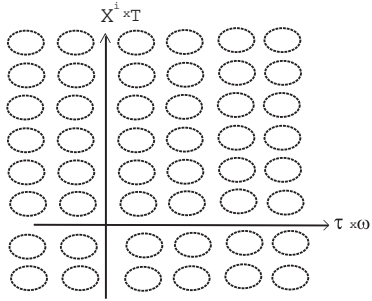}
\end{center}
\label{foam}
\end{figure}

With the momentum cut-off parameter $\La$ for the UV-regularization, 
and $\mu$ for the IR-regularization, 
the condition, (\ref{oneHO7}) or (\ref{NHO10}), can be rewritten as
%*** conc1 %%%%%%%%%%%%%%%%
\bea
\frac{1}{d\tau^2}dX^2 < \frac{\om^2}{T^2}=\frac{\La^2}{\mu^2}\sim\infty \q\mbox{or}\q
\frac{1}{\sqrt{d\tau^2}}\sqrt{dX^2} < \frac{\om}{T}=\frac{\La}{\mu}\sim\infty 
\com
\label{conc1}
\eea 
%%%%%%%%%%%%%%%%%%%%%%%%%%%
where the 'IR-UV harmonic relation' $\mu=\La T/\om$ is used.
\footnote{
The relation appears in the regularization process of 
the numerical evaluation of (\ref{intro2})\cite{SI0812}. 
The choice was taken in Ref.\cite{RS01} for the $\be$-function 
calculation of the 5D warped YM theory. They regarded 
the parameters $\om$ and $T$ as "physical" UV and IR cutoffs respectively.
} 
The above relation looks like a sort of the {\it uncertainty} relation. 
It is the relation between 4D coordinates and the extra one 
, not between coordinates and momenta which makes the phase space.
\footnote{
In the development of the string theory, the uncertainty relation 
between the coordinates is shown to appear\cite{Yoneya87}. 
} 
We need the constraint, expressed by (\ref{conc1}), on the bulk space coordinates 
when we regard the hypersurface as the elastic system. 
If we write the relation (\ref{conc1}) as 
%*** conc2 %%%%%%%%%%%%%%%%
\bea
dX^2~<~\Hcal^2 d\tau^2\q\mbox{or}\q
\sqrt{dX^2}~<~\Hcal \sqrt{d\tau^2}\com\q \Hcal\equiv \om/T 
\com
\label{conc2}
\eea 
%%%%%%%%%%%%%%%%%%%%%%%%%%%
it says the 
length relation between the "smallest" interval in the X-direction and 
that in the $\tau$-direction. 
Here we define the {\it dimensionless constant} $\Hcal$ by the ratio of 
$\om$ and $T$. 
If we can take the assumption: 
%*** conc3 %%%%%%%%%%%%%%%%
\bea
\sqrt{T^2dX^2}\sim \al_X\com\q \sqrt{\om^2d\tau^2}\sim \al_\tau\com\q
\al_X<\al_\tau
\com
\label{conc3}
\eea 
%%%%%%%%%%%%%%%%%%%%%%%%%%%
where $\al_X$ and $\al_\tau$ are regarded dimensionless constants of the order $O(1)$, 
it suggests the foam-like structure\cite{Haw79} of 5D bulk space. 
See Fig.\ref{foam}. 
                             %%%   <Fig.7   %%%
                              %%%   Fig.7>  %%%
The size of the one 'bubble' is about $T^{-1}\times \om^{-1}$. Mathematically 
it suggests a new algebra among the bulk space coordinates.

%%%%%%%%%%%%%%%%%%%%%%%%%%%%  Sec.AppA  %%%%%%%%%%%%%%%%%%%%%%%%%%%%%%%%%
%%%%                                                               %%%%%%
%%%%  General Quantum Statistical System of N Coordinates %%%%%%
%%%%                                                               %%%%%%
%%%%                                                               %%%%%%
%%%%%%%%%%%%%%%%%%%%%%%%%%%%%%%%%%%%%%%%%%%%%%%%%%%%%%%%%%%%%%%%%%%%%%%%%
\section{Appendix A~:~General Quantum Statistical System of N Coordinates\label{GenNHO}}
%***  GenNHO\ \ 
We consider the system of N coordinates, $\{x^1,x^2,\cdots x^N \}$, in the general isotropic potential. 
This is the generalization of Sec.\ref{NHO} where the elastic-type potential is only considered. 
\subsection{'Dirac' Type\label{GenNHOa}}
%***  GenNHOa\q
Let us consider N+1 dim Euclidean space $(X^i,\tau), i=1,2,\cdots ,N$ described by the following metric. 
%*** GenNHO1%%%%%%%%%%%%%%%%
\bea
ds^2=\sum_{i=1}^N(dX^i)^2+2V(r) d\tau^2 =G_{AB}dX^A dX^B\com\nn
A,B=1,2,\cdots,N,N+1;\q X^{N+1}\equiv \tau\com \nn
(G_{AB})=\mbox{diag}(1,1,\cdots,1,2V(r))\com\q
r^2\equiv \sum_{i=1}^N(X^i)^2\com\q (X^A)=(X^i,\tau)\com
\label{GenNHO1}
\eea 
%%%%%%%%%%%%%%%%%%%%%%%%%%%
where the isotropy property in N dim space $\{X^i|i=1\sim N\}$ is assumed. 
The general potential $V$ depends only on $r$ . 
The present convention is given by
%*** GenNHO2%%%%%%%%%%%%%%%%
\bea
\Ga^A_{BC}=\half G^{AD}(\pl_BG_{DC}+\pl_CG_{DB}-\pl_DG_{BC})\com\nn
R^C_{~D,AB}=\pl_A\Ga^C_{BD}+\Ga^C_{EA}\Ga^E_{DB}- A\change B\com\nn
R_{AB}(=R_{BA})=R^C_{~A,BC}
=\pl_B\Ga^C_{CA}-\pl_C\Ga^C_{AB}+\Ga^C_{DB}\Ga^D_{AC}-\Ga^C_{DC}\Ga^D_{AB}\com\q
R=G^{AB}R_{AB}
\pr
\label{GenNHO2}
\eea
%%%%%%%%%%%%%%%%%%%%%%%%%%%
(Sec.\ref{GQSS} is the $N=1$ case. Sec.\ref{NHO} is the case of the potential: $V(r)=\om^2r^2/2$) 
%%%%%%%%%%%%%%%%%%%%%%%%%%%
The explicit result for (\ref{GenNHO1}) is
%*** GenNHO3%%%%%%%%%%%%%%%%
\bea
R_{ij}=\frac{V'}{2rV}\del_{ij}
+\{ \frac{V''}{2V}-\frac{V'}{2rV}-\frac{1}{4}(\frac{V'}{V})^2 \}\frac{X^iX^j}{r^2}
\com\nn 
R_{\tau i}=0\com\q R_{i\tau}=0\com\q 
R_{\tau\tau}=V''-\frac{{V'}^2}{2V}+(N-1)\frac{1}{r}V'\com\nn
R=\frac{V''}{V}-\half(\frac{V'}{V})^2+(N-1)\frac{V'}{rV}\com\q 
\sqrt{G}=\sqrt{2V}
\com
\label{GenNHO3}
\eea 
%%%%%%%%%%%%%%%%%%%%%%%%%%%
where $V'=\frac{dV}{dr}, V''=\frac{d^2V}{dr^2}$. 
The elastic system (of N 'particles') is obtained by setting $V=\om^2r^2/2$.  
%*** GenNHO4 %%%%%%%%%%%%%%%%
\bea
R_{ij}=\frac{\del_{ij}}{r^2}-\frac{X^iX^j}{(r^2)^2}\com\q R_{\tau i}=0\com\q
R_{i\tau}=0\com\q R_{\tau\tau}=(N-1)\om^2\com\nn
R=\frac{2(N-1)}{r^2}\com\q \sqrt{G}R=2(N-1)\frac{\om}{r}
\pr
\label{GenNHO4}
\eea 
%%%%%%%%%%%%%%%%%%%%%%%%%%%
%N=2 case is given in the text (\ref{NHO1b}). 

We impose the periodicity (\ref{oneHO2})(period: $\be$).
%*** NHO2%%%%%%%%%%%%%%%%
%\bea
%\tau\ra\tau+\be
%\pr
%\label{NHO2}
%\eea 
%%%%%%%%%%%%%%%%%%%%%%%%%%%
Here we take a path $\{X^i=x^i(\tau)|\ 0\leq \tau\leq \be,\ i=1,2,\cdots,N\}$ and 
the {\it induced} metric on the line is given by
%*** GenNHO4b %%%%%%%%%%%%%%%%
\bea
X^i=x^i(\tau)\com\q 
dX^i=\xdot^i d\tau\com\q \xdot^i\equiv\frac{dx^i}{d\tau}\com\q 
0\leq\tau\leq\be\com
\nn
ds^2=\left( \sum_{i=1}^{N}({\xdot}^i)^2+2V(r)\right) d\tau^2
\pr
\label{GenNHO4b}
\eea 
%%%%%%%%%%%%%%%%%%%%%%%%%%%
See Fig.\ref{PathLine}. 
Then the length L of the path $\{ x^i(\tau)\}$ is given by 
%*** GenNHO5  %%%%%%%%%%%%%%%%
\bea
L=\int ds=\int_0^\beta\sqrt{ \left( \sum_{i=1}^N(\xdot^i)^2+2V(r)\right) }~d\tau 
\pr
\label{GenNHO5}
\eea 
%%%%%%%%%%%%%%%%%%%%%%%%%%%
Taking the half of the length ($\half L$) as the system Hamiltonian({\it minimal length principle}), 
the free energy $F$ of the system is given by 
%*** GenNHO6  %%%%%%%%%%%%%%%%
\bea
\e^{-\be F}=(\prod_i\int_{-\infty}^{\infty}d\rho_i)
\int_{\begin{array}{c}x^i(0)=\rho_i\\x^i(\be)=\rho_i\end{array}}
\prod_{\tau,i}\Dcal x^i(\tau)\exp \left[-\half\int_0^\beta
\sqrt{ \left( \sum_{i=1}^N(\xdot^i)^2+2V(r)\right) }~d\tau
                                  \right]
\com
\label{GenNHO6}
\eea 
%%%%%%%%%%%%%%%%%%%%%%%%%%%
where the path-integral is done for all possible paths $\{x^i(\tau);i=1,2,\cdots N\}$ with the indicated b.c.. 
We can regard this as the free energy for the general system of N coordinates isotropically interacting.

%2010.6.19  20:00
Instead of the length $L$, we take another geometric quantity. Let us consider 
the N dim {\it hypersurface} in N+1 dim space (a closed-string configuration), (\ref{NHO6}). 
See Fig.\ref{PathHySurf} for the N=2 case. 
The form of 
$r(\tau)$ describes a path (N dimensional hypersurface in the bulk) which is {\it isotropic} in 
the 'brane' at $\tau$ (the N dim plane 'perpendicularly' standing at $\tau$ of the extra axis, 
not the hypersurface ). 
The {\it induced} metric on the N dim hypersurface is given by
%*** GenNHO7%%%%%%%%%%%%%%%%
\bea
ds^2=\sum_{i,j}(\del_{ij}+\frac{2V(r)}{r^2\rdot^2}x^ix^j)dx^idx^j\equiv
\sum_{i,j}g_{ij}dx^idx^j\com\nn
g_{ij}=\del_{ij}+\frac{2V(r)}{r^2\rdot^2}x^ix^j\com\q r^2=\sum_{i=1}^{N}(x^i)^2\com\q
\det(g_{ij})=1+\frac{2V(r)}{\rdot^2}
\pr
\label{GenNHO7}
\eea 
%%%%%%%%%%%%%%%%%%%%%%%%%%%
This is the metric of a O(N) 
nonlinear system.  
Then the area of the N dim hypersurface is given by 
%*** GenNHO8%%%%%%%%%%%%%%%%
\bea
A_N=\int\sqrt{\det g_{ij}}~d^Nx=\frac{N\pi^{N/2}}{\Ga(\frac{N}{2}+1)}\int\sqrt{\rdot^2+2V(r)}r^{N-1}d\tau
\pr
\label{GenNHO8}
\eea 
%%%%%%%%%%%%%%%%%%%%%%%%%%%
When we take $\half A_N$ as the Hamiltonian ({\it minimal area principle}), 
the free energy $F$ is given by 
%*** GenNHO9%%%%%%%%%%%%%%%%
\bea
\e^{-\be F}=\int_{0}^{\infty}d\rho
\int_{\begin{array}{c}r(0)=\rho\\r(\be)=\rho\end{array}}
\prod_{\tau,i}\Dcal x^i(\tau)\exp \left[
-\half\frac{N\pi^{N/2}}{\Ga(\frac{N}{2}+1)}
\int\sqrt{\rdot^2+2V(r)}~r^{N-1}d\tau
                                  \right]
\pr
\label{GenNHO9}
\eea 
%%%%%%%%%%%%%%%%%%%%%%%%%%%

\subsection{Standard Type\label{GenNHOb}}
%***label***{GenNHOb}\nl
Now we consider another type of N+1 dim Euclidean space $(X^i,\tau);\ i=1,2,\cdots N$ described by the following
line element. 
%*** GenNHO9b%%%%%%%%%%%%%%%%
\bea
ds^2=d\tau^{-2}\{\sum_{i=1}^N(dX^i)^2\}^2+4V(r)^2d\tau^2
+4V(r) \{\sum_{j=1}^N(dX^j)^2\}  \nn
=\frac{1}{d\tau^{2}}\{ 
\sum_{i=1}^N(dX^i)^2+2V(r)d\tau^2
                     \}^2  
\com\q r=\sqrt{\sum_{i=1}^N(X^i)^2}\com
\label{GenNHO9b}
\eea 
%%%%%%%%%%%%%%%%%%%%%%%%%%%]
with the condition (\ref{NHO10}) 
in order to keep all terms of (\ref{GenNHO9b}) in the order of $\ep^2$. 
Again we 
note that, in the above case, we do {\it not} have N+1 dim (bulk) metric. 
We impose the periodicity (\ref{oneHO2}): (period: $\be$).
%*** NHO11%%%%%%%%%%%%%%%%
%\bea
%\tau\ra\tau+\be
%\pr
%\label{NHO11}
%\eea 
%%%%%%%%%%%%%%%%%%%%%%%%%%%

Here we take a path $\{x^i(\tau)|\ 0\leq \tau\leq \be, i=1,2,\cdots,N\}$(Fig.\ref{PathLine}) and 
the {\it induced} metric on the path is given by
%*** GenNHO12%%%%%%%%%%%%%%%%
\bea
X^i=x^i(\tau)\com\q dX^i=\xdot^i d\tau\com\q \xdot^i\equiv\frac{dx^i}{d\tau}\com\q 
0\leq\tau\leq\be\com\nn
ds^2=[\sum_{i=1}^N(\xdot^i)^2+2V(r)]^2 d\tau^2\com\q
r=\sqrt{\sum_{i=1}^N(x^i)^2}\
\pr
\label{GenNHO12}
\eea 
%%%%%%%%%%%%%%%%%%%%%%%%%%%
Then the length L of the path $\{x^i(\tau)\}$ is given by 
%*** GenNHO13%%%%%%%%%%%%%%%%
\bea
L[x^i(\tau)]=\int ds=\int_0^\beta\left( \sum_{i=1}^N(\xdot^i)^2+2V(r)\right) d\tau
\pr
\label{GenNHO13}
\eea 
%%%%%%%%%%%%%%%%%%%%%%%%%%%
Hence, taking $\half L$ as the Hamiltonian ({\it minimal length principle}
), 
the free energy $F$ of the system is given by 
%*** GenNHO14%%%%%%%%%%%%%%%%
\bea
\e^{-\be F}=\left( \prod_i\int_{-\infty}^{\infty}d\rho_i \right)
\int_{\begin{array}{c}x^i(0)=\rho_i\\x^i(\be)=\rho_i\end{array}}
\prod_{i,\tau}\Dcal x^i(\tau)\exp \left[-\half\int_0^\beta\left( \sum_{i=1}^N(\xdot^i)^2+2V(r)\right)d\tau
                                  \right]
\com
\label{GenNHO14}
\eea 
%%%%%%%%%%%%%%%%%%%%%%%%%%%
where the path-integral is done for all possible paths with the indicated b.c.. This is the free energy 
for the general isotropic system of N coordinates.

\subsection{Middle type of O(N) nonlinear system\label{GenNHOc}}
%***label***{GenNHOc}\nl
Instead of (\ref{GenNHO9b}), we can start from a slightly modified metric.   
%*** GenNHO15%%%%%%%%%%%%%%%%
\bea
ds^2=4V(r)^2d\tau^2
+4\ka V(r)\{\sum_{j=1}^N(dX^j)^2\}= 
4V(r)\left\{V(r)d\tau^2
+\ka\sum_{j=1}^N(dX^j)^2\right\} 
 \pr
\label{GenNHO15}
\eea 
%%%%%%%%%%%%%%%%%%%%%%%%%%%
We drop the first term of (\ref{GenNHO9b}), and add 
a free (real) parameter $\ka$ in the third one. 
We stress that, in this case, we need {\it not} the condition of (\ref{NHO10}). 
The line element is the ordinary type and 
we have the bulk metric $G_{AB}$ in this case. 
The Ricci tensor and the scalar curvature are given by 
%*** GenNHO15b%%%%%%%%%%%%%%%%
\bea
(G_{AB})=\mbox{diag}(4\ka V,\ 4\ka V,\ \cdots,\ 4\ka V,\ 4 V^2)\com\q
\sqrt{G}=\sqrt{\det G_{AB}}=(4|\ka| V)^{N/2}\cdot 2 V
\com\nn
R_{ij}=\left\{
\frac{N}{2}\frac{1}{r}(\frac{V'}{rV})'-\frac{N-2}{4}(\frac{V'}{rV})^2  
       \right\}X^i X^j  
+\left\{
\half(\frac{V'}{rV})'r+N\frac{V'}{rV}+\frac{N}{4}r^2(\frac{V'}{rV})^2
 \right\}\del_{ij} \com\nn
R_{\tau~i}=R_{i~\tau}=0\com\q
R_{\tau\tau}=\frac{1}{\ka}\left\{
(\frac{V'}{r})'r+N\frac{V'}{r}+\frac{N-2}{2}\frac{{V'}^2}{V}
\right\}\com\q V'=\frac{d}{dr}V(r)\com\nn
R=\frac{1}{4\ka}\left\{
(N+1)\frac{r}{V^2}(\frac{V'}{r})'+\frac{N^2-3N-2}{4}\frac{{V'}^2}{V^3}
+N(N+1)\frac{V'}{r V^2}
                \right\}
\com\q 
r^2=\sum_{i=1}^N (X^i)^2\com
\label{GenNHO15b}
\eea 
%%%%%%%%%%%%%%%%%%%%%%%%%%%
where $i,j=1,2,\cdots N$ and $X^{N+1}=\tau$. 
\footnote{
$R>0 \q\mbox{for}\q \ka>0\com\q R<0 \q\mbox{for}\q \ka<0$. 
} 
We consider the N dim hypersurface (\ref{NHO6}), or Fig.\ref{PathHySurf}, 
and the {\it induced} metric on it is 
given by 
%*** GenNHO16%%%%%%%%%%%%%%%%
\bea
ds^2=\sum_{i,j=1}^N 4V(r)(\ka\del_{ij}+\frac{V(r)}{r^2\rdot^2}x^ix^j)dx^idx^j\equiv
\sum_{i,j}g_{ij}dx^idx^j\com\nn
g_{ij}=4V(r)(\ka\del_{ij}+\frac{V(r)}{r^2\rdot^2}x^ix^j)
\pr
\label{GenNHO16}
\eea 
%%%%%%%%%%%%%%%%%%%%%%%%%%%
Then the area of this hypersurface is given by 
%*** GenNHO17%%%%%%%%%%%%%%%%
\bea
A_N=\int\sqrt{\det g_{ij}}~d^Nx=
\frac{(2\pi\om^2|\ka|)^{N/2}}{\Ga(\frac{N}{2}+1)}
\int_0^\be V^{N/2}\sqrt{\rdot^2+\frac{V(r)}{|\ka|}}~r^{N-1}d\tau
\pr
\label{GenNHO17}
\eea 
%%%%%%%%%%%%%%%%%%%%%%%%%%%
Taking $\half A_N$ as the Hamiltonian ({\it minimal area principle}), the free energy is given by 
%*** GenNHO18%%%%%%%%%%%%%%%%
\bea
\e^{-\be F}=\int_{0}^{\infty}d\rho
\int_{\begin{array}{c}r(0)=\rho\\r(\be)=\rho\end{array}}
\prod_{\tau,i}\Dcal x^i(\tau)\exp \left[
-\half 
\frac{(2\pi\om^2|\ka|)^{N/2}}{\Ga(\frac{N}{2}+1)}
\int_0^\be V^{N/2}\sqrt{\rdot^2+\frac{V(r)}{|\ka|}}~r^{N-1}d\tau
                                  \right]
\pr
\label{GenNHO18}
\eea 
%%%%%%%%%%%%%%%%%%%%%%%%%%%

\subsection{Modified type\label{GenMod}}
%***label***{GenMod}\nl
The general modified metric is given by
%*** GenMod1%%%%%%%%%%%%%%%%
\bea
ds^2= W(\tau)\left\{
2V(r)d\tau^2+\sum_{j=1}^N(dX^j)^2
            \right\} 
 \com
\label{GenMod1}
\eea 
%%%%%%%%%%%%%%%%%%%%%%%%%%%
where $V(r)$ and $W(\tau)$ are general functions of $r$ and $\tau$ respectively. 
(As pointed out in Sec.4.4, the special case:\ 
$ W(\tau)=\frac{1}{\tau^2},\ V(r)=\half $\ 
is Euclidean AdS$_{N+1}$. )

The Ricci tensor and the scalar curvature are given by 
%*** GenMod1b%%%%%%%%%%%%%%%%
\bea
(G_{AB})=\mbox{diag}(W(\tau),\ W(\tau),\ \cdots,\ W(\tau),\ 2 W(\tau)V(r))\com\q\nn
\sqrt{G}=\sqrt{\det G_{AB}}=\sqrt{2}W(\tau)^{(N+1)/2} V(r)^{1/2}
\com\nn
R_{ij}=\left\{
\frac{1}{2}\frac{1}{r V}(\frac{V'}{r})'-\frac{1}{4}(\frac{V'}{rV})^2  
       \right\}X^i X^j  
    +\left\{  
\frac{1}{4 V}\pl_\tau(\frac{\Wdot}{W})+\half\frac{V'}{r V}
+\frac{N-1}{8}(\frac{\Wdot}{W})^2\frac{1}{V}
     \right\}\del_{ij} \com\nn
R_{\tau~i}=R_{i~\tau}=-\frac{N-1}{4}\frac{V'}{V}\frac{\Wdot}{W}
\frac{X^i}{r}\com\q
R_{\tau\tau}=
\frac{N}{2}\pl_\tau(\frac{\Wdot}{W})+(\frac{V'}{r})'r
+N\frac{V'}{r}-\frac{{V'}^2}{2 V}
            \com\nn 
V'=\frac{d}{dr}V(r)\com\q \Wdot=\frac{dW}{d\tau}\com\q 
r^2=\sum_{i=1}^N (X^i)^2\com\nn
R=\frac{1}{W}\left\{
\frac{r}{V}(\frac{V'}{r})'-\half(\frac{V'}{V})^2
+N\frac{V'}{r V}+\frac{N}{2}\frac{1}{V}\frac{{\ddot W}}{W}
+\frac{N(N-5)}{8}\frac{1}{V}(\frac{\Wdot}{W})^2
                \right\}
\pr
\label{GenMod1b}
\eea 
%%%%%%%%%%%%%%%%%%%%%%%%%%% 

We consider the N dim hypersurface (\ref{NHO6}), or Fig.\ref{PathHySurf}, 
and the {\it induced} metric on it is 
given by 
%*** GenMod2%%%%%%%%%%%%%%%%
\bea
ds^2=W(\tau)\sum_{i,j=1}^N \left(
\del_{ij}+\frac{2V(r)}{r^2\rdot^2}x^ix^j
                           \right)dx^idx^j\equiv
\sum_{i,j}g_{ij}dx^idx^j\com\nn
g_{ij}=W(\tau)\left(\del_{ij}+\frac{2V(r)}{r^2\rdot^2}x^ix^j\right)
\pr
\label{GenMod2}
\eea 
%%%%%%%%%%%%%%%%%%%%%%%%%%%
Then the area of this hypersurface is given by 
%*** GenMod3%%%%%%%%%%%%%%%%
\bea
A_N=\int\sqrt{\det g_{ij}}~d^Nx=
\frac{N\pi^{N/2}}{\Ga(\frac{N}{2}+1)}
\int_0^\be W(\tau)^{N/2}\sqrt{\rdot^2+2 V(r)}~r^{N-1}d\tau
\pr
\label{GenMod3}
\eea 
%%%%%%%%%%%%%%%%%%%%%%%%%%%
Taking $\half A_N$ as the Hamiltonian ({\it minimal area principle}), the free energy is given by 
%*** GenMod4%%%%%%%%%%%%%%%%
\bea
\e^{-\be F}=\int_{0}^{\infty}d\rho
\int_{\begin{array}{c}r(0)=\rho\\r(\be)=\rho\end{array}}
\prod_{\tau,i}\Dcal x^i(\tau)\exp \left[
-\half 
\frac{N\pi^{N/2}}{\Ga(\frac{N}{2}+1)}
\int_0^\be W(\tau)^{N/2}\sqrt{\rdot^2+2 V(r)}~r^{N-1}d\tau
                                  \right]
\pr
\label{GenMod4}
\eea 
%%%%%%%%%%%%%%%%%%%%%%%%%%%

\newpage
%\acknowledgement{
\section{Acknowledgment}
Parts of the content of this work have been already presented at 
the international conference on "Particle Physics, Astrophysics and 
Quantum Field Theory"(08.11.27-29, Nanyang Executive Centre, Singapore)\cite{SI0903Singa}, 
YITP Workshop on "Field Theory and String Theory" (09.7.6-10, Kyoto Univ.,Yukawa Memorial Hall), 
First Mediterranean Conference on Classical and Quantum Gravity 
(09.9.14-18, Kolymbari, Crete, Greece)\cite{SI0909}, 
RIMS-YITP Joint Workshop on 'Duality and Scale in Quantum Science'
(09.11.4-6, Kyoto Univ., Kyoto, Japan), Int. Workshop on "Strong Coupling Gauge 
Theories in LHC Era"(09.12.8-11, Nagoya Univ., Nagoya, Japan)\cite{SI0912}, 
KEK Theory Workshop 2010 (10.3.10-13, KEK, Ibaraki, Japan), 
IPMU Workshop on 'Condensed Matter Physics Meets High Energy Physics'
(10.2.8-12, IPMU, Univ. of Tokyo, Kashiwa, Japan), and 
65th(10.3.20-23,Okayama;10.9.11-14,Kita-Kyusyu) 
Japan Physical Society Meeting. 
The author thanks T. Appelquist (Yale Univ.), 
K. Fujikawa (Nihon Univ.), T. Inagaki (Hiroshima Univ.), S. Iso(KEK)
, K. Kanaya (Univ. of Tsukuba), Y. Kitazawa(KEK), T. Kugo (Kyoto Univ.), 
N. Sakai (Tokyo Woman's Christian Univ.), M. Sakamoto (Kobe Univ.), 
M. Tanabashi(Nagoya Univ.), S. Watamura(Tohoku Univ.) and T. Yoneya (Univ. of Tokyo) 
for useful comments and encouragement on the occasions.

\end{document}